\documentclass[10pt]{elsarticle}

\usepackage{fullpage}
\usepackage{parskip}
\usepackage{amsmath}
\usepackage{amssymb}
\usepackage{enumerate}
\usepackage{graphicx}
\usepackage{psfrag}
\usepackage{caption}
\usepackage{subcaption}
\usepackage{color}
\usepackage{MnSymbol}
\usepackage{tikz}

\usepackage{pgfplots}
  \pgfplotsset{compat=newest}
  \usetikzlibrary{plotmarks}
  \usepackage{grffile}
  \usepackage{amsmath}

\newcommand\solidrule[1][0.5cm]{\rule[0.5ex]{#1}{0.6pt}}
\newcommand\dashedrule{\mbox{%
  \solidrule[1mm]\hspace{0.7mm}\solidrule[1mm]}}
\newcommand{\mytriangle[1]}{\tikz{\draw[black,fill=#1] (0,0) -- (0.15cm,0) -- (0.075cm,0.15cm) -- cycle;}}
\newcommand{\mysquare[1]}{\tikz{\draw[black,fill=#1] (0,0) -- (0.13cm,0) -- (0.13cm,0.13cm) -- (0cm,0.13cm) -- cycle;}}
\newcommand{\mydiamond[1]}{\tikz{\draw[black,fill=#1] (0cm,0) -- (0.08cm, -0.08cm) -- (0.16cm,0) -- (0.08cm,0.08cm) -- cycle;}}
\newcommand{\mycircle[1]}{\tikz{\draw[black,fill=#1] (0,0) circle (.55ex);}}

\biboptions{sort&compress}
\newlength\figH
\newlength\figW

\begin{document}

\begin{frontmatter}
\title{Accurate computation of surface stresses and forces with immersed boundary methods}

\author[ad1]{Andres Goza\corref{cor1}}
\cortext[cor1]{Corresponding author}
\ead{ajgoza@gmail.com}
\author[ad1]{Sebastian Liska}
\author[ad1]{Benjamin Morley \fnref{fn1}}
\fntext[fn1]{Current Address: DAMTP, University of Cambridge, Wilberforce Rd, Cambridge CB3 0WA, United Kingdom} 
\author[ad1]{Tim Colonius}

\address[ad1]{Division of Engineering and Applied Science, California Institute of Technology, Pasadena, CA 91125, USA}

\begin{abstract}
Many immersed boundary methods solve for surface stresses that impose the velocity boundary conditions on an immersed body. These surface stresses may contain spurious oscillations that make them ill-suited for representing the physical surface stresses on the body. Moreover, these inaccurate stresses often lead to unphysical oscillations in the history of integrated surface forces such as the coefficient of lift. While the errors in the surface stresses and forces do not necessarily affect the convergence of the velocity field, it is desirable, especially in fluid-structure interaction problems, to obtain smooth and convergent stress distributions on the surface. To this end, we show that the equation for the surface stresses is an integral equation of the first kind whose ill-posedness is the source of spurious oscillations in the stresses. We also demonstrate that for sufficiently smooth delta functions, the oscillations may be filtered out to obtain physically accurate surface stresses. The filtering is applied as a post-processing procedure, so that the convergence of the velocity field is unaffected. We demonstrate the efficacy of the method by computing stresses and forces that converge to the physical stresses and forces for several test problems.
\end{abstract}


\end{frontmatter}

\section{Introduction}
Immersed boundary (IB) methods are attractive for simulating flows around moving or deforming bodies, due in large part to their ability to treat the immersed body and the flow domain with separate grids. The use of different grids removes the need for remeshing, which is often computationally expensive. The original IB method of Peskin used a smeared delta function to represent the surface stresses that related the flow domain to the immersed body \cite{Peskin:1972ty}. These surface stresses were derived using a specific constitutive law governing the deformation of the immersed body. 

A different set of IB methods derives the surface stresses by requiring that they impose the velocity boundary conditions on the immersed body, rather than by directly linking them to the deformation of the solid \cite{Uhlmann:2005gf,Huang:2009ic,Zhang:2007ez,Yang:2009fo,Ji:2012he,Taira:2007jl,Colonius:2008dr,Le:2008en,griffith}. Because they are derived from the boundary conditions on the immersed body, we refer here to these IB methods as surface velocity-based IB methods. These methods produce surface stresses that are poor representations of the physical surface stresses. A subset of these also produce unphysical oscillations in time traces of surface force quantities such as the coefficients of lift and drag, since they enforce the boundary constraint approximately rather than by explicitly solving a linear system \cite{Uhlmann:2005gf,Huang:2009ic,Zhang:2007ez}. Yang \emph{et al.}\ reduced the unphysical oscillations in these surface force quantities \cite{Yang:2009fo}, but to our knowledge the inaccuracies in the surface stresses have not been addressed. This is probably due to the fact that the velocity field converges in spite of these erroneous surface stresses, so surface velocity-based IB methods may be used without modification for problems where accurate knowledge of the surface stresses is not required. 

However, correct information about surface stresses and forces is important in many applications, such as characterizing the performance of wings and aerodynamic bodies in unsteady flows, understanding and controlling flow separation around bluff bodies, and simulating fully coupled flow-structure-interaction (FSI) problems with deforming bodies. In this work, we characterize and remedy the spurious surface stresses and forces obtained by surface-velocity based IB methods.

We show that, for any choice of smeared delta function, the equation for the surface stresses is an integral equation of the first kind whose ill-posedness leads to an inaccurate representation of the high frequency components of the surface stresses. The error in these high frequency components was also observed by Kallemov \emph{et al.}\ for a six point delta function \cite{griffith}. We demonstrate that there is an inverse relation between the smoothness of the smeared delta function and the amplitude of the high frequency components that are needed to accurately represent the physically correct stress. Thus, when sufficiently smooth delta functions are selected, the high-frequency components that are erroneously amplified when solving the integral equation may be effectively filtered out of the solution without damaging the overall surface stress. By contrast, filtering out the incorrect high frequency components for insufficiently smooth smeared delta functions obscures important physical information.

We develop an efficient filtering technique for penalizing the erroneous high frequency stress components. The filtering procedure is performed as a post-processing step, so the convergence of the velocity field is unaffected. We demonstrate that, for all smeared delta functions considered, the filtered stresses are better approximations to the physical stresses than their unfiltered counterparts. However, because of the aforementioned inverse relationship between the smoothness of the smeared delta function and the magnitude of the high frequency components required to represent the physical stresses, this filtering procedure only provides convergent surface stresses when applied to sufficiently smooth smeared delta functions. These results are illustrated for several problems using the immersed boundary projection method (IBPM) of Colonius and Taira \cite{Colonius:2008dr}.

\section{Demonstrating and resolving inaccurate computation of source terms for a model problem}

The difficulty in solving integral equations of the first kind that arise from surface velocity-based IB methods is illustrated and remedied for a model problem in this section. Section 3 will demonstrate that the same type of integral equation arises from the Navier-Stokes equations. Thus, the same techniques developed here may be used to compute surface stresses and forces that arise in fluid flows.

The model problem considered is the Poisson equation for an unknown function $\psi$ on a 2D square domain $\Omega = \{\textbf{x} =[x,\, y]^T : |x|, |y| \le 1 \}$ with an unknown singular source term $f$ that takes nonzero values on an immersed surface denoted by $\Gamma$:
\begin{equation}
\begin{gathered}
\nabla^2 \psi(\textbf{x}) = - \int_\Gamma f(\boldsymbol\xi(s) ) \delta(\textbf{x} - \boldsymbol\xi(s)) ds \\ 
\psi(\textbf{x}) = \psi^{\partial \Omega}(\textbf{x}), \; \textbf{x} \in \partial \Omega \\
\int_\Omega \psi(\textbf{x}) \delta(\textbf{x} - \boldsymbol\xi(s)) d\textbf{x} = \psi^\Gamma(\boldsymbol\xi(s))
\label{eqn:TE_1}
\end{gathered}
\end{equation}
where $s$ is a variable that parametrizes the IB (e.g., arc length), $\boldsymbol\xi(s)$ is the Lagrangian coordinate of a given point on the IB, $\partial \Omega$ is the boundary of the domain $\Omega$, $\psi^{\partial \Omega}(\textbf{x})$ is a function of prescribed values for $\psi$ on $\partial \Omega$, and $\psi^\Gamma(\boldsymbol\xi(s))$ is a function defined on the immersed body. Note that the delta function $\delta(\textbf{x} - \boldsymbol\xi(s) )$ is used to relate quantities between the immersed surface and the solution domain. An error analysis of numerical solutions to (\ref{eqn:TE_1}) has been performed in the case where $f$ is prescribed \cite{Tornberg04, tornberg2010}. To mirror surface velocity-based IB methods, we leave $f$ as an unknown that is solved by explicitly incorporating the third equation as a boundary constraint.

We take $\Gamma$ to be a circle of radius ${1}/{2}$ centered at $\textbf{x} = 0$, $\psi^{\partial \Omega}(\textbf{x}) = 1 - \frac{1}{2}\log(2|\textbf{x}|)$, and $\psi^{\Gamma}(\boldsymbol\xi) = 1$. The exact solution to (\ref{eqn:TE_1}) is then
\begin{align}
\psi_{ex}(\textbf{x}) &=  
	\begin{cases}
		1, & |\textbf{x} | \le \frac{1}{2} \\
		1 - \frac{1}{2}\log(2|\textbf{x}|) & |\textbf{x} | > \frac{1}{2}
	\end{cases}
	\label{eqn:poiss_ex_u}
	\\
	f_{ex}(\boldsymbol\xi) &= 1 \label{eqn:poiss_ex_f}
\end{align}
Another quantity of interest is $F_{ex} = \int_\Gamma f_{ex}(\boldsymbol\xi(s)) ds = \pi$. This term is analogous to the integrated surface force, which is often of interest for IB flow solvers.

To make IB methods suitable for computation, the Dirac delta function in (\ref{eqn:TE_1}) is replaced with a smeared delta function, $\delta_h(\textbf{x} - \boldsymbol\xi(s,t))$, that is continuous and has nonzero but compact support defined in terms of the grid spacing, $h$, of the discretized domain on which the numerical solution is obtained (see, e.g. \cite{Peskin:2002go}). Thus, the numerical solution for a given grid spacing $h$ has as its corresponding continuous solution
\begin{equation}
\psi(\textbf{x}) = - \int_\Omega  \int_\Gamma f(\boldsymbol\xi(s')) \delta_h(\textbf{x}' - \boldsymbol\xi(s')) G(\textbf{x};\textbf{x}')ds'd\textbf{x}'
\label{eqn:poiss_sol}
\end{equation}
where $G(\textbf{x};\textbf{x}')$ is the Green's function for the Poisson problem evaluated at $\textbf{x}$ due to a source at \textbf{x}$'$, and $\delta_h$ is the (continuous) smeared delta function. The equation (\ref{eqn:poiss_sol}) is written in terms of the unknown source term $f$. To arrive at an equation for this source term, we multiply both sides of (\ref{eqn:poiss_sol}) by $\delta_h(\boldsymbol\xi(s) - \textbf{x})$ and integrate over the domain $\Omega$:
\begin{equation}
\int_\Omega  \int_\Omega  \int_\Gamma  f(\boldsymbol\xi(s')) \delta_h(\textbf{x}' - \boldsymbol\xi(s')) G(\textbf{x};\textbf{x}')   \delta_h(\textbf{x} - \boldsymbol\xi(s))ds' d\textbf{x}'d\textbf{x} = - \psi^\Gamma(\boldsymbol\xi(s))
\label{eqn:source_sol}
\end{equation}
The solution $\psi(\textbf{x})$ is then obtained by substituting the solution $f$ of (\ref{eqn:source_sol}) into (\ref{eqn:poiss_sol}). 

Since $\delta_h$ is continuous for a given grid spacing $h$, the kernel in the integral equation (\ref{eqn:source_sol}) is continuous and has finite support. Thus, the integral operator is compact and formally does not have a bounded inverse \cite{Kress2012}. As a consequence, discretizations of this equation lead to inaccurate surface source terms. To highlight that the difficulty in computing the source term occurs for all smeared delta functions, we will use four different functions that are common in the literature. In all cases, the two-dimensional smeared delta function is defined by the tensor product of two one-dimensional smeared delta functions; \emph{i.e.} $\delta_h(\textbf{x} - \boldsymbol\xi) = \delta_h(x - \xi) \delta_h(y - \eta)$, where $\textbf{x} = [x, y]^T$ and $\boldsymbol\xi = [\xi, \eta]^T$. The four one-dimensional smeared delta functions we consider are given below.
\begin{itemize}
	\item{A 2-point hat function:
		\begin{equation}
			\delta_h^{hat}(r) = \begin{cases} \frac{1}{h}  - \frac{|r|}{h^2}, & |r| \le h \\ 0, & |r| > h \end{cases}
		\label{eqn:hat_ddf}
		\end{equation}
	}
	\item{A 3-point function:
		\begin{equation}
			\delta_h^3 (r) = 
				\begin{cases} 
					\frac{1}{3h} \left(1 + \sqrt{1 -3\left(\frac{r}{h}\right)^2} \right), & |r| \le \frac{h}{2} \\ 
					\frac{1}{6h} \left( 5 - \frac{3|r|}{h} - \sqrt{ 1 - 3\left(1 - \frac{|r|}{h}\right)^2 } \right), & \frac{h}{2} \le |r| \le \frac{3h}{2} \\
					0, & |r| > \frac{3h}{2}
				\end{cases}
				\label{eqn:ddf_3}
		\end{equation}
	}
	\item{A 4 point cosine function:
		\begin{equation}
			\delta_h^{cos}(r) = 
				\begin{cases} 
        				\frac{1}{4h}\left(1 + \cos\left(\frac{\pi r}{2h}\right) \right), & |r| \le 2h\\ 
        				0, & |r| > 2h 
				\end{cases}
		\label{eqn:cos_ddf}
		\end{equation}
	}
	\item{A Gaussian function:
		\begin{equation}
			\delta_h^G(r) = 
				\begin{cases}
					\sqrt{\frac{\pi}{36h^2}} e^\frac{-\pi^2 r^2}{36h^2}, & |r| \le 14h \\
					0, &|r| > 14h
				\end{cases}
    		\label{eqn:gauss_ddf}	
		\end{equation}
		
		A Gaussian function formally has infinite support. The parameters and cut-off used in (\ref{eqn:gauss_ddf}) lead to a truncation error on the order of machine precision. Other parameter choices may be selected to satisfy different error tolerances \cite{tornberg2010}. 
	}
\end{itemize}

To solve the problem numerically, we discretize the system (\ref{eqn:TE_1}) as (after replacing the Dirac delta functions with the smeared delta functions $\delta_h$) 
\begin{gather}
L\psi = -Hf \label{eqn:poiss_disc1} + b_L \\
E\psi  = \psi ^\Gamma \label{eqn:poiss_disc2}
\end{gather}
where the variables $\psi $, $\psi ^\Gamma$, and $f$ are understood to be the spatially discrete versions of their continuous counterparts; $L$ is the discrete Laplacian; $b_L$ is a boundary condition term that arises from discretizing the Laplacian operator; and $H(\cdot)$ and $E(\cdot)$ are discretizations of the operations $\int_\Gamma (\cdot) \delta_h(\textbf{x}-\boldsymbol\xi) ds$ and $\int_A (\cdot) \delta_h(\textbf{x}-\boldsymbol\xi) d\textbf{x}$, respectively. Note that the different choices of smeared delta function change $E$ and $H$.

Equations (\ref{eqn:poiss_disc1}) and (\ref{eqn:poiss_disc2}) may be combined to arrive at an equation for $f$, given by
\begin{equation}
EL^{-1}Hf = -\psi^\Gamma + EL^{-1}b_L
\label{eqn:source_eqn_disc}
\end{equation}
which is a discretization of the integral equation (\ref{eqn:source_sol}). Following Colonius and Taira \cite{Colonius:2008dr}, we construct $E$ and $H$ such that $EL^{-1}H$ is positive definite and symmetric.

The simulation for this problem was performed using a finite difference discretization on a uniform grid, with the standard 5 point stencil used for $L$. The grid spacing of the immersed body was chosen to match that of the solution grid. The numerical solution was obtained on the finite domain $[-1,1]\times[-1,1]$; the boundary conditions for $\psi$ were obtained by the exact solution (\ref{eqn:poiss_ex_u}). In what follows, $n_b$ and $n_g$ are the number of points on the immersed body and the solution domain, respectively.

\begin{figure}[h!]
\setlength{\figH}{0.15\textwidth}
\setlength{\figW}{0.98\textwidth}
\centering
%
%
\begin{tikzpicture}

\begin{axis}[%
width=0.22\figW,
height=\figH,
at={(0\figW,0\figH)},
scale only axis,
xmin=0,
xmax=1,
xlabel={$\frac{\theta}{2\pi}$},
ymin=-5,
ymax=5,
ylabel={$f$},
axis background/.style={fill=white},
title style={font=\bfseries},
title={$\delta_h^{hat}$},
ticklabel style={font=\footnotesize},ylabel style={font=\small},xlabel style={font=\small},title style={font=\footnotesize}
]
\addplot [color=black,dashed,line width=0.5pt,mark size=1.2pt,mark=o,mark options={solid},forget plot]
  table[row sep=crcr]{%
0	0.959205901848604\\
0.0333333333333333	-0.808712631595166\\
0.0666666666666667	1.14104487051903\\
0.1	0.404767448169189\\
0.133333333333333	0.608953720756511\\
0.166666666666667	0.438354056980014\\
0.2	1.32578275360448\\
0.233333333333333	0.234545299669474\\
0.266666666666667	0.234545299671486\\
0.3	1.32578275360293\\
0.333333333333333	0.43835405708812\\
0.366666666666667	0.608953720758659\\
0.4	0.404767448168113\\
0.433333333333333	1.14104487052036\\
0.466666666666667	-0.808712631632523\\
0.5	0.959205901932467\\
0.533333333333333	-0.808712631595224\\
0.566666666666667	1.14104487051907\\
0.6	0.404767448164309\\
0.633333333333333	0.608953720764154\\
0.666666666666667	0.438354057301473\\
0.7	1.32578275360223\\
0.733333333333333	0.234545299653701\\
0.766666666666667	0.234545299809342\\
0.8	1.32578275360914\\
0.833333333333333	0.438354057518065\\
0.866666666666667	0.608953720759217\\
0.9	0.404767448166834\\
0.933333333333333	1.14104487051571\\
0.966666666666667	-0.808712631599558\\
};
\addplot [color=black,solid,line width=0.5pt,forget plot]
  table[row sep=crcr]{%
0	1\\
0.0333333333333333	1\\
0.0666666666666667	1\\
0.1	1\\
0.133333333333333	1\\
0.166666666666667	1\\
0.2	1\\
0.233333333333333	1\\
0.266666666666667	1\\
0.3	1\\
0.333333333333333	1\\
0.366666666666667	1\\
0.4	1\\
0.433333333333333	1\\
0.466666666666667	1\\
0.5	1\\
0.533333333333333	1\\
0.566666666666667	1\\
0.6	1\\
0.633333333333333	1\\
0.666666666666667	1\\
0.7	1\\
0.733333333333333	1\\
0.766666666666667	1\\
0.8	1\\
0.833333333333333	1\\
0.866666666666667	1\\
0.9	1\\
0.933333333333333	1\\
0.966666666666667	1\\
};
\end{axis}

\begin{axis}[%
width=0.22\figW,
height=\figH,
at={(0.24\figW,0\figH)},
scale only axis,
xmin=0,
xmax=1,
xlabel={$\frac{\theta}{2\pi}$},
ymin=-5,
ymax=5,
ytick={-5,-4,-3,-2,-1,0,1,2,3,4,5},
yticklabels={\empty},
axis background/.style={fill=white},
title style={font=\bfseries},
title={$\delta_h^{3}$},
ticklabel style={font=\footnotesize},ylabel style={font=\small},xlabel style={font=\small},title style={font=\footnotesize}
]
\addplot [color=black,dashed,line width=0.5pt,mark size=1.7pt,mark=diamond,mark options={solid},forget plot]
  table[row sep=crcr]{%
0	1.00259466244837\\
0.0333333333333333	0.899239253785942\\
0.0666666666666667	0.913787603572213\\
0.1	1.38147289811983\\
0.133333333333333	1.16888994275984\\
0.166666666666667	1.56855747492027\\
0.2	0.952419477063263\\
0.233333333333333	0.99373869886644\\
0.266666666666667	0.993738699086773\\
0.3	0.952419477067276\\
0.333333333333333	1.56855747490949\\
0.366666666666667	1.1688899427791\\
0.4	1.38147289810364\\
0.433333333333333	0.91378760357355\\
0.466666666666667	0.89923925354239\\
0.5	1.00259466246269\\
0.533333333333333	0.899239253677066\\
0.566666666666667	0.913787603533179\\
0.6	1.38147289808016\\
0.633333333333333	1.16888994275051\\
0.666666666666667	1.56855747488107\\
0.7	0.952419477062871\\
0.733333333333333	0.993738699199842\\
0.766666666666667	0.993738698836685\\
0.8	0.952419477059078\\
0.833333333333333	1.56855747488275\\
0.866666666666667	1.16888994273749\\
0.9	1.38147289810784\\
0.933333333333333	0.913787603562222\\
0.966666666666667	0.899239253760598\\
};
\addplot [color=black,solid,line width=0.5pt,forget plot]
  table[row sep=crcr]{%
0	1\\
0.0333333333333333	1\\
0.0666666666666667	1\\
0.1	1\\
0.133333333333333	1\\
0.166666666666667	1\\
0.2	1\\
0.233333333333333	1\\
0.266666666666667	1\\
0.3	1\\
0.333333333333333	1\\
0.366666666666667	1\\
0.4	1\\
0.433333333333333	1\\
0.466666666666667	1\\
0.5	1\\
0.533333333333333	1\\
0.566666666666667	1\\
0.6	1\\
0.633333333333333	1\\
0.666666666666667	1\\
0.7	1\\
0.733333333333333	1\\
0.766666666666667	1\\
0.8	1\\
0.833333333333333	1\\
0.866666666666667	1\\
0.9	1\\
0.933333333333333	1\\
0.966666666666667	1\\
};
\end{axis}

\begin{axis}[%
width=0.22\figW,
height=\figH,
at={(0.48\figW,0\figH)},
scale only axis,
xmin=0,
xmax=1,
xlabel={$\frac{\theta}{2\pi}$},
ymin=-5,
ymax=5,
ytick={-5,-4,-3,-2,-1,0,1,2,3,4,5},
yticklabels={\empty},
axis background/.style={fill=white},
title style={font=\bfseries},
title={$\delta_h^{cos}$},
ticklabel style={font=\footnotesize},ylabel style={font=\small},xlabel style={font=\small},title style={font=\footnotesize}
]
\addplot [color=black,dashed,line width=0.5pt,mark size=1.6pt,mark=triangle,mark options={solid},forget plot]
  table[row sep=crcr]{%
0	146.130966705494\\
0.0333333333333333	0.618606818668674\\
0.0666666666666667	-2.65961568553747\\
0.1	3.6123691047123\\
0.133333333333333	-0.966442115472091\\
0.166666666666667	-40.2992958340456\\
0.2	2.74945997540204\\
0.233333333333333	-0.36668786169954\\
0.266666666666667	-0.366687864124342\\
0.3	2.74945997716823\\
0.333333333333333	-40.2992958332425\\
0.366666666666667	-0.966442115815013\\
0.4	3.6123691043363\\
0.433333333333333	-2.65961568487214\\
0.466666666666667	0.618606819819877\\
0.5	146.130965134331\\
0.533333333333333	0.618606818802146\\
0.566666666666667	-2.65961568538832\\
0.6	3.61236910429321\\
0.633333333333333	-0.966442115905374\\
0.666666666666667	-40.2992958336567\\
0.7	2.74945997991189\\
0.733333333333333	-0.366687863343435\\
0.766666666666667	-0.366687863757576\\
0.8	2.74945997387492\\
0.833333333333333	-40.2992958342383\\
0.866666666666667	-0.96644211593123\\
0.9	3.61236910474363\\
0.933333333333333	-2.65961568601868\\
0.966666666666667	0.618606819080352\\
};
\addplot [color=black,solid,line width=0.5pt,forget plot]
  table[row sep=crcr]{%
0	1\\
0.0333333333333333	1\\
0.0666666666666667	1\\
0.1	1\\
0.133333333333333	1\\
0.166666666666667	1\\
0.2	1\\
0.233333333333333	1\\
0.266666666666667	1\\
0.3	1\\
0.333333333333333	1\\
0.366666666666667	1\\
0.4	1\\
0.433333333333333	1\\
0.466666666666667	1\\
0.5	1\\
0.533333333333333	1\\
0.566666666666667	1\\
0.6	1\\
0.633333333333333	1\\
0.666666666666667	1\\
0.7	1\\
0.733333333333333	1\\
0.766666666666667	1\\
0.8	1\\
0.833333333333333	1\\
0.866666666666667	1\\
0.9	1\\
0.933333333333333	1\\
0.966666666666667	1\\
};
\end{axis}

\begin{axis}[%
width=0.22\figW,
height=\figH,
at={(0.72\figW,0\figH)},
scale only axis,
xmin=0,
xmax=1,
xlabel={$\frac{\theta}{2\pi}$},
ymin=-5,
ymax=5,
ytick={-5,-4,-3,-2,-1,0,1,2,3,4,5},
yticklabels={\empty},
axis background/.style={fill=white},
title style={font=\bfseries},
title={$\delta_h^{G}$},
ticklabel style={font=\footnotesize},ylabel style={font=\small},xlabel style={font=\small},title style={font=\small}
]
\addplot [color=black,dashed,line width=0.5pt,mark size=1.1pt,mark=square,mark options={solid},forget plot]
  table[row sep=crcr]{%
0	0.993333944971108\\
0.0333333333333333	1.34435280600799\\
0.0666666666666667	1.14403357066892\\
0.1	0.931732176949928\\
0.133333333333333	1.00169935263775\\
0.166666666666667	0.584442768945512\\
0.2	0.852656292800911\\
0.233333333333333	0.949879427710357\\
0.266666666666667	0.949768473886007\\
0.3	0.852651043879322\\
0.333333333333333	0.584457917852172\\
0.366666666666667	1.00162016755533\\
0.4	0.931853253284808\\
0.433333333333333	1.14403671984796\\
0.466666666666667	1.34433331977604\\
0.5	0.993258458372648\\
0.533333333333333	1.3443178267366\\
0.566666666666667	1.14394421676693\\
0.6	0.93170344041655\\
0.633333333333333	1.00179067002604\\
0.666666666666667	0.584333581891089\\
0.7	0.852543630336253\\
0.733333333333333	0.949600439578683\\
0.766666666666667	0.949794132449133\\
0.8	0.852537197693469\\
0.833333333333333	0.584504686669461\\
0.866666666666667	1.00173289304374\\
0.9	0.931660534121386\\
0.933333333333333	1.14391093621321\\
0.966666666666667	1.34440834173902\\
};
\addplot [color=black,solid,line width=0.5pt,forget plot]
  table[row sep=crcr]{%
0	1\\
0.0333333333333333	1\\
0.0666666666666667	1\\
0.1	1\\
0.133333333333333	1\\
0.166666666666667	1\\
0.2	1\\
0.233333333333333	1\\
0.266666666666667	1\\
0.3	1\\
0.333333333333333	1\\
0.366666666666667	1\\
0.4	1\\
0.433333333333333	1\\
0.466666666666667	1\\
0.5	1\\
0.533333333333333	1\\
0.566666666666667	1\\
0.6	1\\
0.633333333333333	1\\
0.666666666666667	1\\
0.7	1\\
0.733333333333333	1\\
0.766666666666667	1\\
0.8	1\\
0.833333333333333	1\\
0.866666666666667	1\\
0.9	1\\
0.933333333333333	1\\
0.966666666666667	1\\
};
\end{axis}
\end{tikzpicture}%
    \caption[waste]{Computed source term ($f$) versus arc length along the cylinder for the Poisson model problem; \protect\solidrule: $f_{ex}$. All plots used $h = 1/640$.}
    \label{fig:source_plot}
\end{figure}
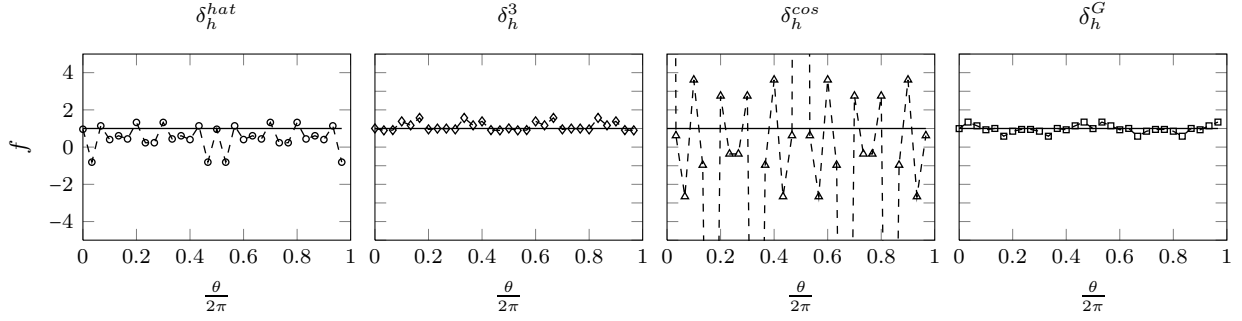

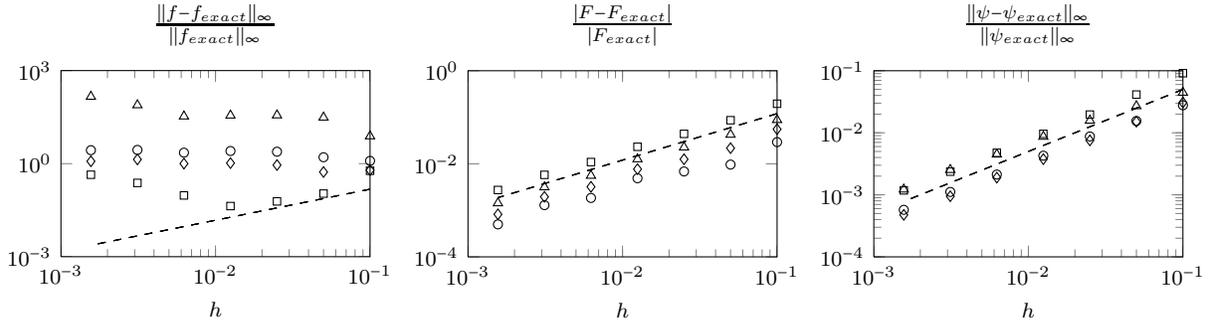
\begin{figure}[h!]
\setlength{\figH}{0.15\textwidth}
\setlength{\figW}{0.95\textwidth}
\centering
%
%
\begin{tikzpicture}

\begin{axis}[%
width=0.262\figW,
height=\figH,
at={(0.689\figW,0\figH)},
scale only axis,
xmode=log,
xmin=0.001,
xmax=0.1,
xminorticks=true,
xlabel={$h$},
ymode=log,
ymin=0.0001,
ymax=0.1,
yminorticks=true,
axis background/.style={fill=white},
title style={font=\bfseries},
title={$\frac{||\psi - \psi_{exact}||_\infty}{||\psi_{exact}||_\infty}$},
ticklabel style={font=\footnotesize},ylabel style={font=\small},xlabel style={font=\footnotesize},title style={font=\normalsize}
]
\addplot [color=black,mark size=1.7pt,only marks,mark=o,mark options={solid},forget plot]
  table[row sep=crcr]{%
0.1	0.02775749794882\\
0.05	0.0155054897269182\\
0.025	0.0087232035221726\\
0.0125	0.00427569998244415\\
0.00625	0.00212359099916171\\
0.003125	0.00111381306800573\\
0.0015625	0.000573265757480401\\
};
\addplot [color=black,dashed,forget plot]
  table[row sep=crcr]{%
0.1	0.05\\
0.05	0.025\\
0.025	0.0125\\
0.0125	0.00625\\
0.00625	0.003125\\
0.003125	0.0015625\\
0.0015625	0.00078125\\
};
\addplot [color=black,mark size=2.0pt,only marks,mark=triangle,mark options={solid},forget plot]
  table[row sep=crcr]{%
0.1	0.0445283371838503\\
0.05	0.0269764580791968\\
0.025	0.0157806349454586\\
0.0125	0.00867715853484041\\
0.00625	0.00449137216499895\\
0.003125	0.00252845853900818\\
0.0015625	0.00122465318424492\\
};
\addplot [color=black,dashed,forget plot]
  table[row sep=crcr]{%
0.1	0.05\\
0.05	0.025\\
0.025	0.0125\\
0.0125	0.00625\\
0.00625	0.003125\\
0.003125	0.0015625\\
0.0015625	0.00078125\\
};
\addplot [color=black,mark size=2.0pt,only marks,mark=diamond,mark options={solid},forget plot]
  table[row sep=crcr]{%
0.1	0.0307557795430764\\
0.05	0.015099006256116\\
0.025	0.00753750773472039\\
0.0125	0.00376688781587387\\
0.00625	0.00189503387814494\\
0.003125	0.000949201571304403\\
0.0015625	0.000475255339821956\\
};
\addplot [color=black,dashed,forget plot]
  table[row sep=crcr]{%
0.1	0.05\\
0.05	0.025\\
0.025	0.0125\\
0.0125	0.00625\\
0.00625	0.003125\\
0.003125	0.0015625\\
0.0015625	0.00078125\\
};
\addplot [color=black,mark size=1.4pt,only marks,mark=square,mark options={solid},forget plot]
  table[row sep=crcr]{%
0.1	0.0912642043713501\\
0.05	0.0412939347330212\\
0.025	0.0196534445614971\\
0.0125	0.00960703577659117\\
0.00625	0.0047532421588492\\
0.003125	0.00236479684957591\\
0.0015625	0.00117955174786766\\
};
\addplot [color=black,dashed,forget plot]
  table[row sep=crcr]{%
0.1	0.05\\
0.05	0.025\\
0.025	0.0125\\
0.0125	0.00625\\
0.00625	0.003125\\
0.003125	0.0015625\\
0.0015625	0.00078125\\
};
\end{axis}

\begin{axis}[%
width=0.262\figW,
height=\figH,
at={(0\figW,0\figH)},
scale only axis,
xmode=log,
xmin=0.001,
xmax=0.1,
xminorticks=true,
xlabel={$h$},
ymode=log,
ymin=0.001,
ymax=1000,
yminorticks=true,
axis background/.style={fill=white},
title style={font=\bfseries},
title={$\frac{||f - f_{exact}||_\infty}{||f_{exact}||_\infty}$},
ticklabel style={font=\footnotesize},ylabel style={font=\small},xlabel style={font=\footnotesize},title style={font=\normalsize}
]
\addplot [color=black,mark size=1.7pt,only marks,mark=o,mark options={solid},forget plot]
  table[row sep=crcr]{%
0.1	1.23082556634068\\
0.05	1.6073367747605\\
0.025	2.42764280399111\\
0.0125	2.55612292420957\\
0.00625	2.26166825377009\\
0.003125	2.77058482325304\\
0.0015625	2.71695865058521\\
};
\addplot [color=black,dashed,forget plot]
  table[row sep=crcr]{%
0.1	0.15\\
0.05	0.075\\
0.025	0.0375\\
0.0125	0.01875\\
0.00625	0.009375\\
0.003125	0.0046875\\
0.0015625	0.00234375\\
};
\addplot [color=black,mark size=2.0pt,only marks,mark=triangle,mark options={solid},forget plot]
  table[row sep=crcr]{%
0.1	7.69055750360359\\
0.05	30.9084377737279\\
0.025	36.3970694114873\\
0.0125	35.9614257867926\\
0.00625	33.5875116957267\\
0.003125	77.196369706073\\
0.0015625	145.130966705494\\
};
\addplot [color=black,dashed,forget plot]
  table[row sep=crcr]{%
0.1	0.15\\
0.05	0.075\\
0.025	0.0375\\
0.0125	0.01875\\
0.00625	0.009375\\
0.003125	0.0046875\\
0.0015625	0.00234375\\
};
\addplot [color=black,mark size=2.0pt,only marks,mark=diamond,mark options={solid},forget plot]
  table[row sep=crcr]{%
0.1	0.61632545617172\\
0.05	0.536986230100077\\
0.025	0.9015704813548\\
0.0125	1.04306827309446\\
0.00625	1.0023574853051\\
0.003125	1.36676469045611\\
0.0015625	1.19324403095158\\
};
\addplot [color=black,dashed,forget plot]
  table[row sep=crcr]{%
0.1	0.15\\
0.05	0.075\\
0.025	0.0375\\
0.0125	0.01875\\
0.00625	0.009375\\
0.003125	0.0046875\\
0.0015625	0.00234375\\
};
\addplot [color=black,mark size=1.4pt,only marks,mark=square,mark options={solid},forget plot]
  table[row sep=crcr]{%
0.1	0.589590533606347\\
0.05	0.108490319020626\\
0.025	0.0607716688270157\\
0.0125	0.0429714110307839\\
0.00625	0.0946272468737575\\
0.003125	0.239244537682005\\
0.0015625	0.438913573983706\\
};
\addplot [color=black,dashed,forget plot]
  table[row sep=crcr]{%
0.1	0.15\\
0.05	0.075\\
0.025	0.0375\\
0.0125	0.01875\\
0.00625	0.009375\\
0.003125	0.0046875\\
0.0015625	0.00234375\\
};
\end{axis}

\begin{axis}[%
width=0.262\figW,
height=\figH,
at={(0.345\figW,0\figH)},
scale only axis,
xmode=log,
xmin=0.001,
xmax=0.1,
xminorticks=true,
xlabel={$h$},
ymode=log,
ymin=0.0001,
ymax=1,
ytick={0.0001,   0.01,      1},
yminorticks=true,
axis background/.style={fill=white},
title style={font=\bfseries},
title={$\frac{|F - F_{exact}|}{|F_{exact}|}$},
ticklabel style={font=\footnotesize},ylabel style={font=\small},xlabel style={font=\footnotesize},title style={font=\normalsize}
]
\addplot [color=black,mark size=1.7pt,only marks,mark=o,mark options={solid},forget plot]
  table[row sep=crcr]{%
0.1	0.0292864337850124\\
0.05	0.00961980792910781\\
0.025	0.00685714855050082\\
0.0125	0.00489382373336593\\
0.00625	0.00184765913888067\\
0.003125	0.00129398247242556\\
0.0015625	0.00049958145042208\\
};
\addplot [color=black,dashed,forget plot]
  table[row sep=crcr]{%
0.1	0.12\\
0.05	0.06\\
0.025	0.03\\
0.0125	0.015\\
0.00625	0.0075\\
0.003125	0.00375\\
0.0015625	0.001875\\
};
\addplot [color=black,mark size=2.0pt,only marks,mark=triangle,mark options={solid},forget plot]
  table[row sep=crcr]{%
0.1	0.0879493681283761\\
0.05	0.0424525735529025\\
0.025	0.0226363214952547\\
0.0125	0.0126173229068684\\
0.00625	0.00557420295796195\\
0.003125	0.00314980431001316\\
0.0015625	0.001429632521851\\
};
\addplot [color=black,dashed,forget plot]
  table[row sep=crcr]{%
0.1	0.12\\
0.05	0.06\\
0.025	0.03\\
0.0125	0.015\\
0.00625	0.0075\\
0.003125	0.00375\\
0.0015625	0.001875\\
};
\addplot [color=black,mark size=2.0pt,only marks,mark=diamond,mark options={solid},forget plot]
  table[row sep=crcr]{%
0.1	0.0553797399886466\\
0.05	0.0217119617613845\\
0.025	0.0125411094669684\\
0.0125	0.00768886095410811\\
0.00625	0.00320521191934034\\
0.003125	0.00195095868581468\\
0.0015625	0.000821741370620905\\
};
\addplot [color=black,dashed,forget plot]
  table[row sep=crcr]{%
0.1	0.12\\
0.05	0.06\\
0.025	0.03\\
0.0125	0.015\\
0.00625	0.0075\\
0.003125	0.00375\\
0.0015625	0.001875\\
};
\addplot [color=black,mark size=1.4pt,only marks,mark=square,mark options={solid},forget plot]
  table[row sep=crcr]{%
0.1	0.195481547921311\\
0.05	0.0864383844309421\\
0.025	0.043994659297322\\
0.0125	0.0232146607456629\\
0.00625	0.0109038032977631\\
0.003125	0.00578860233674422\\
0.0015625	0.00273715134287646\\
};
\addplot [color=black,dashed,forget plot]
  table[row sep=crcr]{%
0.1	0.12\\
0.05	0.06\\
0.025	0.03\\
0.0125	0.015\\
0.00625	0.0075\\
0.003125	0.00375\\
0.0015625	0.001875\\
};
\end{axis}
\end{tikzpicture}%
\caption[waste]{Errors in $f$, $F$, and $\psi$ versus grid spacing ($h$) for the Poisson model problem. \protect\mycircle[none]\,: $\delta_h^{hat}$,  \protect\mydiamond[none]\,: $\delta_h^{3}$, \protect\mytriangle[none]\,: $\delta_h^{cos}$, \protect\mysquare[none]\,: $\delta_h^{G}$, \protect\dashedrule\,: first order convergence.}
\label{fig:err_poiss}
\end{figure}

Figure \ref{fig:source_plot} shows that regardless of the choice of smeared delta function, the source term from (\ref{eqn:source_eqn_disc}) contains spurious oscillations. Moreover, Figure \ref{fig:err_poiss} demonstrates that these inaccuracies persist as the grid is refined, so that $f$ does not converge to $f_{ex}$ as the grid spacing is decreased. Despite this lack of convergence in $f$, the integrated source term $F$ and solution $\psi$ converge at first order to $F_{ex}$ and $\psi_{ex}$, respectively (see Figure \ref{fig:err_poiss}). Convergence of $F$ is a feature of solving (\ref{eqn:source_eqn_disc}); methods that enforce the boundary constraint approximately contain inaccuracies in $F$ as well as $f$ \cite{Uhlmann:2005gf,Huang:2009ic,Zhang:2007ez}, though these were improved by Yang \emph{et al.}\ \cite{Yang:2009fo}. When used with sufficiently smooth smeared delta functions, the method we propose at the end of this section produces convergent approximations for both.

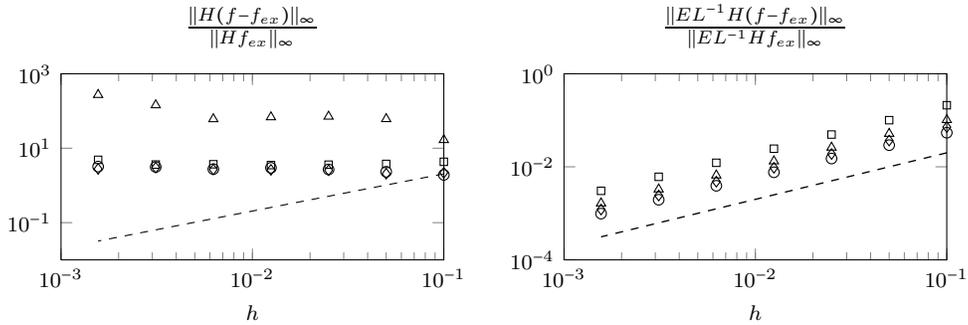
\begin{figure}[h!]
\setlength{\figH}{0.15\textwidth}
\setlength{\figW}{0.75\textwidth}
\centering
%
%
\begin{tikzpicture}

\begin{axis}[%
width=0.411\figW,
height=\figH,
at={(0.54\figW,0\figH)},
scale only axis,
xmode=log,
xmin=0.001,
xmax=0.1,
xminorticks=true,
xlabel={$h$},
ymode=log,
ymin=0.0001,
ymax=1,
ytick={0.0001,   0.01,      1},
yminorticks=true,
axis background/.style={fill=white},
title style={font=\bfseries},
title={$\frac{||EL^{-1} H (f - f_{ex}) ||_\infty}{||EL^{-1} H f_{ex}||_\infty}$},
ticklabel style={font=\footnotesize},ylabel style={font=\footnotesize},xlabel style={font=\footnotesize},title style={font=\normalsize}
]
\addplot [color=black,only marks,mark=o,mark options={solid},forget plot]
  table[row sep=crcr]{%
0.1	0.0540296122964578\\
0.05	0.0291726790068018\\
0.025	0.0150474932195278\\
0.0125	0.0076279908598359\\
0.00625	0.00391255686770606\\
0.003125	0.00195978084771792\\
0.0015625	0.000985867288570701\\
};
\addplot [color=black,only marks,mark=diamond,mark options={solid},forget plot]
  table[row sep=crcr]{%
0.1	0.0713472458252971\\
0.05	0.0366858196942657\\
0.025	0.0188545389788241\\
0.0125	0.00953531082667246\\
0.00625	0.00480317462004321\\
0.003125	0.00241399045921814\\
0.0015625	0.00120761759002419\\
};
\addplot [color=black,only marks,mark=triangle,mark options={solid},forget plot]
  table[row sep=crcr]{%
0.1	0.101791006561671\\
0.05	0.0515427000482543\\
0.025	0.0260329926465431\\
0.0125	0.0132122107276835\\
0.00625	0.0065712553872184\\
0.003125	0.00329293360028418\\
0.0015625	0.0016501095374815\\
};
\addplot [color=black,only marks,mark=square,mark size=1.4pt,mark options={solid},forget plot]
  table[row sep=crcr]{%
0.1	0.210688336701814\\
0.05	0.100528939230368\\
0.025	0.0493482476070603\\
0.0125	0.0244950617691592\\
0.00625	0.0122076120526769\\
0.003125	0.00609444572047391\\
0.0015625	0.00304496158830105\\
};
\addplot [color=black,dashed,line width=0.5pt,forget plot]
  table[row sep=crcr]{%
0.1	0.02\\
0.05	0.01\\
0.025	0.005\\
0.0125	0.0025\\
0.00625	0.00125\\
0.003125	0.000625\\
0.0015625	0.0003125\\
};
\end{axis}

\begin{axis}[%
width=0.411\figW,
height=\figH,
at={(0\figW,0\figH)},
scale only axis,
xmode=log,
xmin=0.001,
xmax=0.1,
xminorticks=true,
xlabel={$h$},
ymode=log,
ymin=0.01,
ymax=1000,
yminorticks=true,
axis background/.style={fill=white},
title style={font=\bfseries},
title={$\frac{||H (f - f_{ex}) ||_\infty}{||H f_{ex} ||_\infty}$},
ticklabel style={font=\footnotesize},ylabel style={font=\footnotesize},xlabel style={font=\footnotesize},title style={font=\normalsize}
]
\addplot [color=black,only marks,mark=o,mark options={solid},forget plot]
  table[row sep=crcr]{%
0.1	1.92154647865198\\
0.05	2.3289437919995\\
0.025	2.68568030949537\\
0.0125	2.97400450769826\\
0.00625	2.74544691310322\\
0.003125	3.12914533190095\\
0.0015625	3.20897857424668\\
};
\addplot [color=black,only marks,mark=diamond,mark options={solid},forget plot]
  table[row sep=crcr]{%
0.1	2.18602631436535\\
0.05	2.0607477454684\\
0.025	2.561347484588\\
0.0125	2.61960824227701\\
0.00625	2.68667457724019\\
0.003125	3.00810278710578\\
0.0015625	2.7794369704854\\
};
\addplot [color=black,only marks,mark=triangle,mark options={solid},forget plot]
  table[row sep=crcr]{%
0.1	16.696417767404\\
0.05	61.0820653871102\\
0.025	71.2037500087665\\
0.0125	68.4568113153477\\
0.00625	60.690792335081\\
0.003125	143.843177749197\\
0.0015625	272.030220575077\\
};
\addplot [color=black,only marks,mark=square,mark size=1.4pt,mark options={solid},forget plot]
  table[row sep=crcr]{%
0.1	4.3122104820837\\
0.05	3.78417364393075\\
0.025	3.60664953451144\\
0.0125	3.51679556156772\\
0.00625	3.70212033365969\\
0.003125	3.64115184619114\\
0.0015625	4.87240388953195\\
};
\addplot [color=black,dashed,forget plot]
  table[row sep=crcr]{%
0.1	2.05\\
0.05	1.025\\
0.025	0.5125\\
0.0125	0.25625\\
0.00625	0.128125\\
0.003125	0.0640625\\
0.0015625	0.03203125\\
};
\end{axis}
\end{tikzpicture}%
\caption[waste]{Errors in $Hf$ and $EL^{-1}Hf$ versus grid spacing ($h$) for the Poisson model problem. \protect\mycircle[none]\,: $\delta_h^{hat}$,  \protect\mydiamond[none]\,: $\delta_h^{3}$, \protect\mytriangle[none]\,: $\delta_h^{cos}$, \protect\mysquare[none]\,: $\delta_h^{G}$, \protect\dashedrule\,: first order convergence.}
\label{fig:err_ELinvHf}
\end{figure}

As shown in Figure \ref{fig:err_ELinvHf}, $f$ has the property that $Hf$ does not converge to $Hf_{ex}$ but $EL^{-1}Hf$ converges to $EL^{-1}Hf_{ex}$. By virtue of (\ref{eqn:source_eqn_disc}), the convergence of $EL^{-1}Hf$ is a statement that using the exact force, $f_{ex}$, to enforce the boundary condition would lead to a boundary value that is not equal to $\psi^\Gamma$ but that converges at first order. This intuitive result was also shown by Tornberg and Engquist \cite{Tornberg04}, and will be exploited in what follows to compute accurate approximations to $f_{ex}$. 

To better explain the results of Figure \ref{fig:err_ELinvHf}, we compute the singular value decomposition (SVD) of $EL^{-1}$. Let $EL^{-1} = U\Sigma V^{T}$, where $U\in\mathbb{R}^{n_b\times n_b}$ and $V\in\mathbb{R}^{n_g \times n_b}$ are matrices of left and right orthonormal singular vectors of $EL^{-1}$, respectively; and $\Sigma\in\mathbb{R}^{n_b \times n_b}$ is a diagonal matrix containing the positive singular values of $EL^{-1}$. The singular values $\sigma_1,\dots,\sigma_{n_b}$ are arranged such that $\sigma_1\ge \sigma_2\ge \dots \ge \sigma_{n_b} >0$, and the singular vectors are notated such that $u_j$ ($v_j$) is the left (right) singular vector corresponding to $\sigma_j$. 

Using this decomposition, $Hf_{ex}$ may be written as a projection onto the basis of vectors formed by $V$:
\begin{equation}
Hf_{ex} = \sum_{j = 1}^{n_b} \alpha_j^{ex} v_j
\label{eqn:Hfproj}
\end{equation}
and $EL^{-1}Hf_{ex}$ may be expressed as
\begin{equation}
EL^{-1}Hf _{ex}= \sum_{j=1}^{n_b} \alpha_j^{ex} \sigma_j u_j 
\label{eqn:ELinvHf}
\end{equation}
where $\alpha_j^{ex} := (v_j^THf_{ex})$. Analogous expressions exist for $Hf$ by replacing $f_{ex}$ with $f$ in (\ref{eqn:Hfproj}) and (\ref{eqn:ELinvHf}). We denote the corresponding coefficients as $\alpha_j := (v_j^THf)$. 

\begin{figure}[h!]
\setlength{\figH}{0.2\textwidth}
\setlength{\figW}{0.95\textwidth}
\centering
     \input{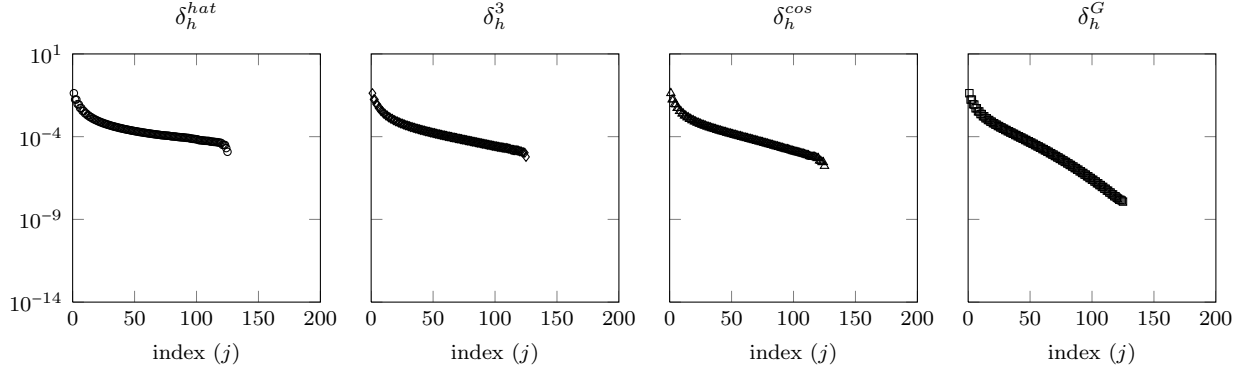}
\caption[blah]{Singular values $\sigma_j$ of $EL^{-1}$ versus index ($j$) for the Poisson model problem. A grid spacing of $h = 1/80$ was used.}
\label{fig:svaldecay}
\end{figure}

Using (\ref{eqn:Hfproj}) and (\ref{eqn:ELinvHf}), Figures \ref{fig:err_ELinvHf} (a) and (b) show that the sum $\sum_{j=1}^{n_b}\alpha_j$ does not converge to $\sum_{j=1}^{n_b}\alpha_j^{ex}$ under grid refinement, but does converge when scaled by the $\sigma_j$. Since $EL^{-1}$ is a discrete integral operator, the $\sigma_j$ decay to very small values \cite{Hansen:1998tf} (see Figure \ref{fig:svaldecay}). Thus, the error in the sum $\sum_{j=1}^{n_b} \alpha_j$ stems from the high index coefficients $\alpha_j$ corresponding to the small $\sigma_j$. The key to computing accurate source terms is to use a smeared delta function for which the coefficients $\alpha_j^{ex}$ decay as rapidly as possible. The spurious high index coefficients $\alpha_j$ may then be filtered out to obtain physical source terms. By contrast, it is difficult to accurately compute source terms using smeared delta functions for which the $\alpha_j^{ex}$ decay slowly, because the incorrect high index $\alpha_j$ obscure important physical information.

Since $EL^{-1}$ is a discrete integral operator, the basis vectors $v_j$ are closely related to the standard Fourier basis \cite{Hansen:1998tf}, and (\ref{eqn:Hfproj}) behaves like an expansion of $Hf_{ex}$ in this basis. The decay rate of the coefficients $\alpha_j^{ex}$ is therefore governed by the smoothness of $Hf_{ex}$, which is determined by the smoothness of the smeared delta function. This is true because $Hf_{ex}$ is a discretization of $\int_\Omega f_{ex}(\boldsymbol\xi) \delta_h(\textbf{x}-\boldsymbol\xi) d\textbf{x} $, and
\begin{equation}
\frac{ d}{d\textbf{x}} \int_\Omega f_{ex}(\boldsymbol\xi) \delta_h(\textbf{x}-\boldsymbol\xi) d\textbf{x} = \int_\Omega f_{ex}(\boldsymbol\xi)  \frac{ d}{d\textbf{x}} \delta_h(\textbf{x}-\boldsymbol\xi) d\textbf{x} 
\end{equation} 

\begin{figure}[h!]
\setlength{\figH}{0.2\textwidth}
\setlength{\figW}{0.95\textwidth}
\centering
     \input{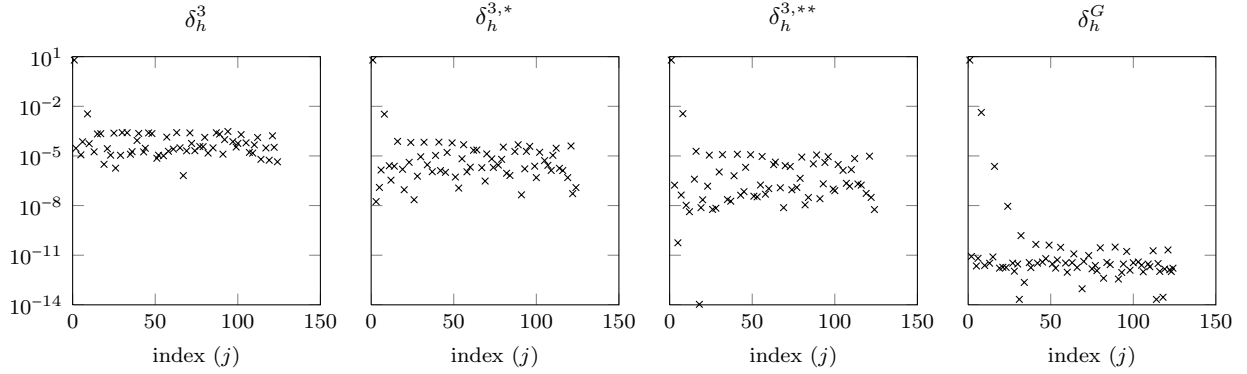}
\caption{Coefficients $\alpha_j^{ex}$ for successively smooth smeared delta functions. Obtained using $h = 1/80$. Note the log scale on the $y$-axis.}
\label{fig:coeff_decay_poiss}
\end{figure}

To demonstrate the effect of the smoothness of the smeared delta function on the decay rate of the coefficients $\alpha_j^{ex}$, we consider a sequence of successively smoother delta functions using the recursive formula developed by Yang \emph{et al.}\ \cite{Yang:2009fo}. Define the operator $\mathcal{S}$ acting on a function $g(r)$ by
\begin{equation}
\mathcal{S}[g(r)] = \int_{r-1/2}^{r+1/2} g(\tilde{r}) d\tilde{r}
\end{equation}
Then the functions we consider are $\delta_h^{3,*}(r) = \mathcal{S}[\delta_h^3(r)]$, $\delta_h^{3,**} = \mathcal{S}[\delta_h^{3,*}(r)]$, and $\delta_h^G$, which as a Gaussian may roughly be thought of as the limit of applying $\mathcal{S}$ to $\delta_h^3$ infinitely many times. Note that $\delta_h^3\in C^1$, $\delta_h^{3,*} \in C^2$, $\delta_h^{3,**} \in C^3$, and $\delta_h^G\in C^\infty$. Figure \ref{fig:coeff_decay_poiss} shows that the decay rate of the coefficients $\alpha_j^{ex}$ increases as smoothness of the smeared delta function increases (note the log scale of the $y$-axis). Note that the compactness of a function in Fourier space is roughly inversely related to its compactness in physical space (see, e.g. \cite{tornberg2010}), so it is important to pick smeared delta functions whose support is not too narrow. 

We now discuss the efficient filtering of the spurious high index coefficients $\alpha_j$. One may in principle filter out the high index coefficients using the SVD of $EL^{-1}$, but this is a costly procedure. Instead, we penalize the spurious components of $f$ by pre-multiplying it with the matrix $\tilde{E}H$, where $\tilde{E} = EW$ is a weighted interpolant that takes the smeared source term $Hf$ back onto the immersed body while preserving its integral value. The filtered source term is then $\tilde{f} = \tilde{E}Hf$. To give the specific form for $W$, define $\boldsymbol{1} = [1, 1, \cdots, 1]^T\in \mathbb{R}^{n_g \times 1}$ and let $(H\boldsymbol{1})_i$ be the $i^{th}$ entry in the  vector $H\boldsymbol{1}$. Then $W$ is a diagonal matrix with entries given by 
\begin{equation}
W_{ii} =  
\begin{cases}
1/(H\boldsymbol{1})_i, & (H\boldsymbol{1})_i \ne 0 \\
0, & \text{else}
\end{cases}
\end{equation} 
Note that $W$ only applies a nonzero weight if the grid point is within the support of the smeared delta function.

The filter $\tilde{E}H$ redistributes the source term $f$ by convolving it with a kernel of smeared delta functions. The weighting matrix leads to a kernel of the same form as is used in nonparametric kernel smoothing techniques \cite{epanechnikov}, and was inspired from work in this field. As shown below, $\tilde{E}H$ filters the high index coefficients at a rate proportional to the smoothness of the smeared delta function being used. This is due to the fact that $\tilde{E}H$ is itself an integral operator, and therefore the decay rate of its singular values is governed by the smoothness of its kernel \cite{Hansen:1998tf}.

\begin{figure}[h!]
\setlength{\figH}{0.3\textwidth}
\setlength{\figW}{0.98\textwidth}
\centering
     \input{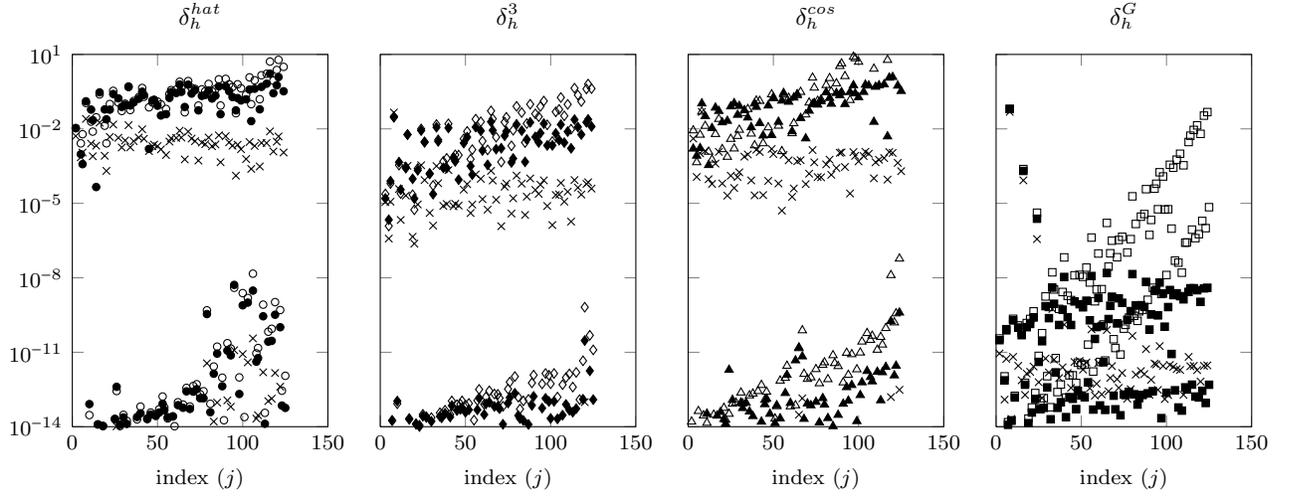}
    \caption[waste]{Coefficients $\alpha_j^{ex}$ ($\times$), $\alpha_j$ (open markers) and $\tilde{\alpha}_j$ (filled markers) for the Poisson model problem. Note the log scale on the $y$-axis. The grid spacing $h = 1/80$ was used.}
    \label{fig:coeff_decay_EHtilde}
\end{figure}

Figure \ref{fig:coeff_decay_EHtilde} demonstrates the effect of filtering by showing the coefficients $\alpha_j^{ex}$, $\alpha_j$ and $\tilde{\alpha}_j:=(v_j^TH\tilde{f})$. Consistent with the observations made above, the high index coefficients $\alpha_j$ are substantially different from those of $\alpha_j^{ex}$. For all smeared delta functions, the filtered coefficients are better approximations to the exact coefficients. Noting that $\delta_h^{hat}\in C^0$, $\delta_h^3 \in C^1$, $\delta_h^{cos}\in C^0$, and $\delta_h^G \in C^\infty$, it is clear from Figure \ref{fig:coeff_decay_EHtilde} that the absolute error in the high frequency $\tilde{\alpha}_j$ decreases as the smoothness of the smeared delta function increases. This is because the magnitude of the high index coefficients $\alpha_j^{ex}$ is smaller for smoother smeared delta functions, so the spurious high index $\alpha_j$ may be filtered more aggressively.

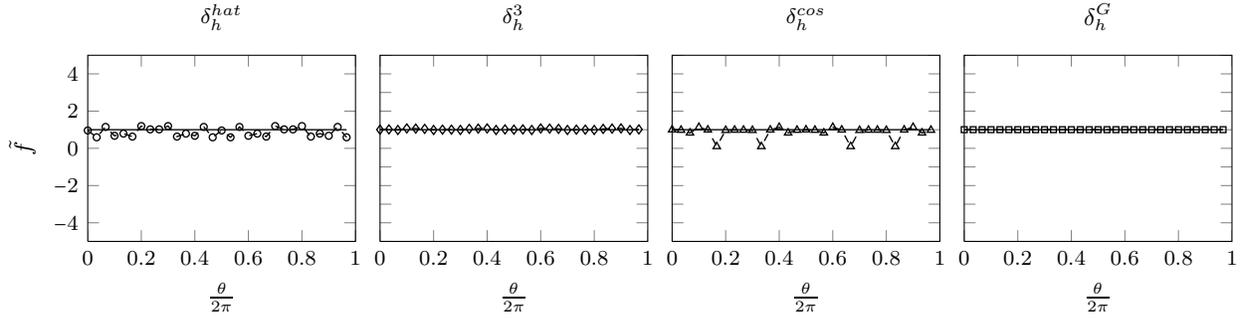
\begin{figure}[h!]
\setlength{\figH}{0.15\textwidth}
\setlength{\figW}{0.98\textwidth}
\centering
%
%
\begin{tikzpicture}

\begin{axis}[%
width=0.22\figW,
height=\figH,
at={(0\figW,0\figH)},
scale only axis,
xmin=0,
xmax=1,
xlabel={$\frac{\theta}{2\pi}$},
ymin=-5,
ymax=5,
ylabel={$\tilde{f}$},
axis background/.style={fill=white},
title style={font=\bfseries},
title={$\delta_h^{hat}$},
ticklabel style={font=\footnotesize},ylabel style={font=\small},xlabel style={font=\small},title style={font=\footnotesize}
]
\addplot [color=black,dashed,line width=0.5pt,mark size=1.3pt,mark=o,mark options={solid},forget plot]
  table[row sep=crcr]{%
0	0.959205901848604\\
0.0333333333333333	0.59224436286659\\
0.0666666666666667	1.14963669608286\\
0.1	0.674282273216386\\
0.133333333333333	0.791607949159339\\
0.166666666666667	0.630610413501891\\
0.2	1.19592866175068\\
0.233333333333333	1.01946009255647\\
0.266666666666667	1.01946009255428\\
0.3	1.19592866174952\\
0.333333333333333	0.630610413581683\\
0.366666666666667	0.79160794915979\\
0.4	0.674282273215658\\
0.433333333333333	1.14963669608339\\
0.466666666666667	0.592244362862827\\
0.5	0.95920590193243\\
0.533333333333333	0.592244362864343\\
0.566666666666667	1.14963669608286\\
0.6	0.674282273214363\\
0.633333333333333	0.791607949161786\\
0.666666666666667	0.630610413739571\\
0.7	1.19592866174927\\
0.733333333333333	1.01946009255774\\
0.766666666666667	1.01946009256078\\
0.8	1.19592866175328\\
0.833333333333333	0.63061041389971\\
0.866666666666667	0.791607949159505\\
0.9	0.674282273215021\\
0.933333333333333	1.14963669608061\\
0.966666666666667	0.592244362867025\\
};
\addplot [color=black,solid,line width=0.5pt,forget plot]
  table[row sep=crcr]{%
0	1\\
0.0333333333333333	1\\
0.0666666666666667	1\\
0.1	1\\
0.133333333333333	1\\
0.166666666666667	1\\
0.2	1\\
0.233333333333333	1\\
0.266666666666667	1\\
0.3	1\\
0.333333333333333	1\\
0.366666666666667	1\\
0.4	1\\
0.433333333333333	1\\
0.466666666666667	1\\
0.5	1\\
0.533333333333333	1\\
0.566666666666667	1\\
0.6	1\\
0.633333333333333	1\\
0.666666666666667	1\\
0.7	1\\
0.733333333333333	1\\
0.766666666666667	1\\
0.8	1\\
0.833333333333333	1\\
0.866666666666667	1\\
0.9	1\\
0.933333333333333	1\\
0.966666666666667	1\\
};
\end{axis}

\begin{axis}[%
width=0.22\figW,
height=\figH,
at={(0.24\figW,0\figH)},
scale only axis,
xmin=0,
xmax=1,
xlabel={$\frac{\theta}{2\pi}$},
ymin=-5,
ymax=5,
ytick={-5,-4,-3,-2,-1,0,1,2,3,4,5},
yticklabels={\empty},
axis background/.style={fill=white},
title style={font=\bfseries},
title={$\delta_h^{3}$},
ticklabel style={font=\footnotesize},ylabel style={font=\small},xlabel style={font=\small},title style={font=\footnotesize}
]
\addplot [color=black,dashed,line width=0.5pt,mark size=1.7pt,mark=diamond,mark options={solid},forget plot]
  table[row sep=crcr]{%
0	1.00261310649465\\
0.0333333333333333	1.01714438659812\\
0.0666666666666667	0.984977404029142\\
0.1	1.06866802168028\\
0.133333333333333	1.06006257792983\\
0.166666666666667	1.04521627907787\\
0.2	0.986419743452977\\
0.233333333333333	1.00403372241065\\
0.266666666666667	1.00403372241268\\
0.3	0.986419743453732\\
0.333333333333333	1.04521627907672\\
0.366666666666667	1.06006257793188\\
0.4	1.06866802167951\\
0.433333333333333	0.984977404029563\\
0.466666666666667	1.01714438658778\\
0.5	1.00261310649681\\
0.533333333333333	1.01714438659111\\
0.566666666666667	0.984977404025759\\
0.6	1.06866802167683\\
0.633333333333333	1.06006257793006\\
0.666666666666667	1.04521627907401\\
0.7	0.986419743452339\\
0.733333333333333	1.00403372241628\\
0.766666666666667	1.00403372241145\\
0.8	0.986419743452962\\
0.833333333333333	1.04521627907399\\
0.866666666666667	1.06006257792761\\
0.9	1.06866802167976\\
0.933333333333333	0.984977404027775\\
0.966666666666667	1.01714438659791\\
};
\addplot [color=black,solid,line width=0.5pt,forget plot]
  table[row sep=crcr]{%
0	1\\
0.0333333333333333	1\\
0.0666666666666667	1\\
0.1	1\\
0.133333333333333	1\\
0.166666666666667	1\\
0.2	1\\
0.233333333333333	1\\
0.266666666666667	1\\
0.3	1\\
0.333333333333333	1\\
0.366666666666667	1\\
0.4	1\\
0.433333333333333	1\\
0.466666666666667	1\\
0.5	1\\
0.533333333333333	1\\
0.566666666666667	1\\
0.6	1\\
0.633333333333333	1\\
0.666666666666667	1\\
0.7	1\\
0.733333333333333	1\\
0.766666666666667	1\\
0.8	1\\
0.833333333333333	1\\
0.866666666666667	1\\
0.9	1\\
0.933333333333333	1\\
0.966666666666667	1\\
};
\end{axis}

\begin{axis}[%
width=0.22\figW,
height=\figH,
at={(0.48\figW,0\figH)},
scale only axis,
xmin=0,
xmax=1,
xlabel={$\frac{\theta}{2\pi}$},
ymin=-5,
ymax=5,
ytick={-5,-4,-3,-2,-1,0,1,2,3,4,5},
yticklabels={\empty},
axis background/.style={fill=white},
title style={font=\bfseries},
title={$\delta_h^{cos}$},
ticklabel style={font=\footnotesize},ylabel style={font=\small},xlabel style={font=\small},title style={font=\footnotesize}
]
\addplot [color=black,dashed,line width=0.5pt,mark size=1.6pt,mark=triangle,mark options={solid},forget plot]
  table[row sep=crcr]{%
0	0.997773125304581\\
0.0333333333333333	0.977488548802927\\
0.0666666666666667	0.830672330336858\\
0.1	1.13137696248985\\
0.133333333333333	0.990975544671748\\
0.166666666666667	0.100220033455738\\
0.2	0.968868552510635\\
0.233333333333333	0.981724617776945\\
0.266666666666667	0.981724617777628\\
0.3	0.96886855252648\\
0.333333333333333	0.100220033476306\\
0.366666666666667	0.99097554466355\\
0.4	1.13137696247587\\
0.433333333333333	0.830672330345186\\
0.466666666666667	0.977488548806001\\
0.5	0.997773125302277\\
0.533333333333333	0.977488548802778\\
0.566666666666667	0.83067233034012\\
0.6	1.13137696248243\\
0.633333333333333	0.990975544655011\\
0.666666666666667	0.100220033462488\\
0.7	0.968868552551331\\
0.733333333333333	0.981724617776707\\
0.766666666666667	0.981724617777199\\
0.8	0.968868552502125\\
0.833333333333333	0.100220033451428\\
0.866666666666667	0.990975544659433\\
0.9	1.13137696249036\\
0.933333333333333	0.830672330329612\\
0.966666666666667	0.97748854880457\\
};
\addplot [color=black,solid,line width=0.5pt,forget plot]
  table[row sep=crcr]{%
0	1\\
0.0333333333333333	1\\
0.0666666666666667	1\\
0.1	1\\
0.133333333333333	1\\
0.166666666666667	1\\
0.2	1\\
0.233333333333333	1\\
0.266666666666667	1\\
0.3	1\\
0.333333333333333	1\\
0.366666666666667	1\\
0.4	1\\
0.433333333333333	1\\
0.466666666666667	1\\
0.5	1\\
0.533333333333333	1\\
0.566666666666667	1\\
0.6	1\\
0.633333333333333	1\\
0.666666666666667	1\\
0.7	1\\
0.733333333333333	1\\
0.766666666666667	1\\
0.8	1\\
0.833333333333333	1\\
0.866666666666667	1\\
0.9	1\\
0.933333333333333	1\\
0.966666666666667	1\\
};
\end{axis}

\begin{axis}[%
width=0.22\figW,
height=\figH,
at={(0.72\figW,0\figH)},
scale only axis,
xmin=0,
xmax=1,
xlabel={$\frac{\theta}{2\pi}$},
ymin=-5,
ymax=5,
ytick={-5,-4,-3,-2,-1,0,1,2,3,4,5},
yticklabels={\empty},
axis background/.style={fill=white},
title style={font=\bfseries},
title={$\delta_h^{G}$},
ticklabel style={font=\footnotesize},ylabel style={font=\small},xlabel style={font=\small},title style={font=\small}
]
\addplot [color=black,dashed,line width=0.5pt,mark size=1.1pt,mark=square,mark options={solid},forget plot]
  table[row sep=crcr]{%
0	1.00305267517844\\
0.0333333333333333	1.00305078539721\\
0.0666666666666667	1.00304572064957\\
0.1	1.00304037702606\\
0.133333333333333	1.00303904124977\\
0.166666666666667	1.00304278207763\\
0.2	1.00304852745383\\
0.233333333333333	1.00305218270407\\
0.266666666666667	1.00305218269601\\
0.3	1.00304852745056\\
0.333333333333333	1.00304278207387\\
0.366666666666667	1.00303904124611\\
0.4	1.00304037703292\\
0.433333333333333	1.00304572065615\\
0.466666666666667	1.00305078539718\\
0.5	1.00305267517001\\
0.533333333333333	1.00305078539679\\
0.566666666666667	1.00304572064621\\
0.6	1.00304037702379\\
0.633333333333333	1.00303904125178\\
0.666666666666667	1.00304278207075\\
0.7	1.00304852744717\\
0.733333333333333	1.00305218270159\\
0.766666666666667	1.0030521827053\\
0.8	1.00304852745657\\
0.833333333333333	1.0030427820873\\
0.866666666666667	1.00303904125583\\
0.9	1.00304037702249\\
0.933333333333333	1.00304572065388\\
0.966666666666667	1.00305078540488\\
};
\addplot [color=black,solid,line width=0.5pt,forget plot]
  table[row sep=crcr]{%
0	1\\
0.0333333333333333	1\\
0.0666666666666667	1\\
0.1	1\\
0.133333333333333	1\\
0.166666666666667	1\\
0.2	1\\
0.233333333333333	1\\
0.266666666666667	1\\
0.3	1\\
0.333333333333333	1\\
0.366666666666667	1\\
0.4	1\\
0.433333333333333	1\\
0.466666666666667	1\\
0.5	1\\
0.533333333333333	1\\
0.566666666666667	1\\
0.6	1\\
0.633333333333333	1\\
0.666666666666667	1\\
0.7	1\\
0.733333333333333	1\\
0.766666666666667	1\\
0.8	1\\
0.833333333333333	1\\
0.866666666666667	1\\
0.9	1\\
0.933333333333333	1\\
0.966666666666667	1\\
};
\end{axis}
\end{tikzpicture}%
    \caption[waste]{$\tilde{f}$ vs arc length along the cylinder for the Poisson problem. The exact solution $f_{ex}$ is given by the solid line (\protect\solidrule) for reference.}
    \label{fig:source_fil_plot}
\end{figure}

\begin{figure}[h!]
\setlength{\figH}{0.15\textwidth}
\setlength{\figW}{0.95\textwidth}
\centering
%
%
\begin{tikzpicture}

\begin{axis}[%
width=0.262\figW,
height=\figH,
at={(0.689\figW,0\figH)},
scale only axis,
xmode=log,
xmin=0.001,
xmax=0.1,
xminorticks=true,
xlabel={$h$},
ymode=log,
ymin=0.0001,
ymax=0.1,
yminorticks=true,
axis background/.style={fill=white},
title style={font=\bfseries},
title={$\frac{||\psi - \psi_{exact}||_\infty}{||\psi_{exact}||_\infty}$},
ticklabel style={font=\footnotesize},ylabel style={font=\small},xlabel style={font=\footnotesize},title style={font=\normalsize}
]
\addplot [color=black,mark size=1.7pt,only marks,mark=o,mark options={solid},forget plot]
  table[row sep=crcr]{%
0.1	0.02775749794882\\
0.05	0.0155054897269182\\
0.025	0.0087232035221726\\
0.0125	0.00427569998244415\\
0.00625	0.00212359099916171\\
0.003125	0.00111381306800573\\
0.0015625	0.000573265757480401\\
};
\addplot [color=black,dashed,forget plot]
  table[row sep=crcr]{%
0.1	0.05\\
0.05	0.025\\
0.025	0.0125\\
0.0125	0.00625\\
0.00625	0.003125\\
0.003125	0.0015625\\
0.0015625	0.00078125\\
};
\addplot [color=black,mark size=2.0pt,only marks,mark=diamond,mark options={solid},forget plot]
  table[row sep=crcr]{%
0.1	0.0307557795430764\\
0.05	0.015099006256116\\
0.025	0.00753750773472039\\
0.0125	0.00376688781587387\\
0.00625	0.00189503387814494\\
0.003125	0.000949201571304403\\
0.0015625	0.000475255339821956\\
};
\addplot [color=black,dashed,forget plot]
  table[row sep=crcr]{%
0.1	0.05\\
0.05	0.025\\
0.025	0.0125\\
0.0125	0.00625\\
0.00625	0.003125\\
0.003125	0.0015625\\
0.0015625	0.00078125\\
};
\addplot [color=black,mark size=2.0pt,only marks,mark=triangle,mark options={solid},forget plot]
  table[row sep=crcr]{%
0.1	0.0445283371838503\\
0.05	0.0269764580791968\\
0.025	0.0157806349454586\\
0.0125	0.00867715853484041\\
0.00625	0.00449137216499895\\
0.003125	0.00252845853900818\\
0.0015625	0.00122465318424492\\
};
\addplot [color=black,dashed,forget plot]
  table[row sep=crcr]{%
0.1	0.05\\
0.05	0.025\\
0.025	0.0125\\
0.0125	0.00625\\
0.00625	0.003125\\
0.003125	0.0015625\\
0.0015625	0.00078125\\
};
\addplot [color=black,mark size=1.4pt,only marks,mark=square,mark options={solid},forget plot]
  table[row sep=crcr]{%
0.1	0.0912642043713501\\
0.05	0.0412939347330212\\
0.025	0.0196534445614971\\
0.0125	0.00960703577659117\\
0.00625	0.0047532421588492\\
0.003125	0.00236479684957591\\
0.0015625	0.00117955174786766\\
};
\addplot [color=black,dashed,forget plot]
  table[row sep=crcr]{%
0.1	0.05\\
0.05	0.025\\
0.025	0.0125\\
0.0125	0.00625\\
0.00625	0.003125\\
0.003125	0.0015625\\
0.0015625	0.00078125\\
};
\end{axis}

\begin{axis}[%
width=0.262\figW,
height=\figH,
at={(0\figW,0\figH)},
scale only axis,
xmode=log,
xmin=0.001,
xmax=0.1,
xminorticks=true,
xlabel={$h$},
ymode=log,
ymin=0.001,
ymax=1,
yminorticks=true,
axis background/.style={fill=white},
title style={font=\bfseries},
title={$\frac{||\tilde{f} - f_{exact}||_\infty}{||f_{exact}||_\infty}$},
ticklabel style={font=\footnotesize},ylabel style={font=\small},xlabel style={font=\footnotesize},title style={font=\normalsize}
]
\addplot [color=black,mark size=1.7pt,only marks,mark=o,mark options={solid},forget plot]
  table[row sep=crcr]{%
0.1	0.548871377833118\\
0.05	0.576672177650452\\
0.025	0.708012795245414\\
0.0125	0.714474474670684\\
0.00625	0.663617809215562\\
0.003125	0.85686168790997\\
0.0015625	0.747943469865711\\
};
\addplot [color=black,dashed,forget plot]
  table[row sep=crcr]{%
0.1	0.15\\
0.05	0.075\\
0.025	0.0375\\
0.0125	0.01875\\
0.00625	0.009375\\
0.003125	0.0046875\\
0.0015625	0.00234375\\
};
\addplot [color=black,mark size=2.0pt,only marks,mark=diamond,mark options={solid},forget plot]
  table[row sep=crcr]{%
0.1	0.145467323440435\\
0.05	0.106555750598228\\
0.025	0.118216463310621\\
0.0125	0.137849641697788\\
0.00625	0.143075280987556\\
0.003125	0.184511163613083\\
0.0015625	0.165298628664354\\
};
\addplot [color=black,dashed,forget plot]
  table[row sep=crcr]{%
0.1	0.15\\
0.05	0.075\\
0.025	0.0375\\
0.0125	0.01875\\
0.00625	0.009375\\
0.003125	0.0046875\\
0.0015625	0.00234375\\
};
\addplot [color=black,mark size=2.0pt,only marks,mark=triangle,mark options={solid},forget plot]
  table[row sep=crcr]{%
0.1	0.30016984201707\\
0.05	0.552311349289287\\
0.025	0.66172876697033\\
0.0125	0.71486555886244\\
0.00625	0.809698290570897\\
0.003125	0.869778875847323\\
0.0015625	0.899779966548572\\
};
\addplot [color=black,dashed,forget plot]
  table[row sep=crcr]{%
0.1	0.15\\
0.05	0.075\\
0.025	0.0375\\
0.0125	0.01875\\
0.00625	0.009375\\
0.003125	0.0046875\\
0.0015625	0.00234375\\
};
\addplot [color=black,mark size=1.4pt,only marks,mark=square,mark options={solid},forget plot]
  table[row sep=crcr]{%
0.1	0.227838834578385\\
0.05	0.106639614035361\\
0.025	0.0510417116918078\\
0.0125	0.0249275896244288\\
0.00625	0.0123177658770983\\
0.003125	0.00612305000290991\\
0.0015625	0.00305267691468103\\
};
\addplot [color=black,dashed,forget plot]
  table[row sep=crcr]{%
0.1	0.15\\
0.05	0.075\\
0.025	0.0375\\
0.0125	0.01875\\
0.00625	0.009375\\
0.003125	0.0046875\\
0.0015625	0.00234375\\
};
\end{axis}

\begin{axis}[%
width=0.262\figW,
height=\figH,
at={(0.345\figW,0\figH)},
scale only axis,
xmode=log,
xmin=0.001,
xmax=0.1,
xminorticks=true,
xlabel={$h$},
ymode=log,
ymin=0.0001,
ymax=1,
ytick={0.0001,   0.01,      1},
yminorticks=true,
axis background/.style={fill=white},
title style={font=\bfseries},
title={$\frac{|F - F_{exact}|}{|F_{exact}|}$},
ticklabel style={font=\footnotesize},ylabel style={font=\small},xlabel style={font=\footnotesize},title style={font=\normalsize}
]
\addplot [color=black,mark size=1.7pt,only marks,mark=o,mark options={solid},forget plot]
  table[row sep=crcr]{%
0.1	0.0292864337850127\\
0.05	0.00961980792910767\\
0.025	0.00685714855050167\\
0.0125	0.00489382373336649\\
0.00625	0.00184765913888067\\
0.003125	0.001293982472425\\
0.0015625	0.000499581450421514\\
};
\addplot [color=black,dashed,forget plot]
  table[row sep=crcr]{%
0.1	0.12\\
0.05	0.06\\
0.025	0.03\\
0.0125	0.015\\
0.00625	0.0075\\
0.003125	0.00375\\
0.0015625	0.001875\\
};
\addplot [color=black,mark size=2.0pt,only marks,mark=diamond,mark options={solid},forget plot]
  table[row sep=crcr]{%
0.1	0.0553797399886465\\
0.05	0.0217119617613842\\
0.025	0.0125411094669684\\
0.0125	0.00768886095410769\\
0.00625	0.00320521191934034\\
0.003125	0.0019509586858144\\
0.0015625	0.000821741370620764\\
};
\addplot [color=black,dashed,forget plot]
  table[row sep=crcr]{%
0.1	0.12\\
0.05	0.06\\
0.025	0.03\\
0.0125	0.015\\
0.00625	0.0075\\
0.003125	0.00375\\
0.0015625	0.001875\\
};
\addplot [color=black,mark size=2.0pt,only marks,mark=triangle,mark options={solid},forget plot]
  table[row sep=crcr]{%
0.1	0.0879493681283761\\
0.05	0.0424525735529024\\
0.025	0.022636321495255\\
0.0125	0.0126173229068693\\
0.00625	0.00557420295796322\\
0.003125	0.00314980431002574\\
0.0015625	0.00142963252185737\\
};
\addplot [color=black,dashed,forget plot]
  table[row sep=crcr]{%
0.1	0.12\\
0.05	0.06\\
0.025	0.03\\
0.0125	0.015\\
0.00625	0.0075\\
0.003125	0.00375\\
0.0015625	0.001875\\
};
\addplot [color=black,mark size=1.4pt,only marks,mark=square,mark options={solid},forget plot]
  table[row sep=crcr]{%
0.1	0.195474023267891\\
0.05	0.0864383844189503\\
0.025	0.0439946592855802\\
0.0125	0.0232146607336956\\
0.00625	0.0109038032863541\\
0.003125	0.00578860232522496\\
0.0015625	0.00273715133132143\\
};
\addplot [color=black,dashed,forget plot]
  table[row sep=crcr]{%
0.1	0.12\\
0.05	0.06\\
0.025	0.03\\
0.0125	0.015\\
0.00625	0.0075\\
0.003125	0.00375\\
0.0015625	0.001875\\
};
\end{axis}
\end{tikzpicture}%
\caption[waste]{Errors in $\tilde{f}$, $F$, and $\psi$ versus grid spacing ($h$) for the Poisson problem. \protect\mycircle[none]\,: $\delta_h^{hat}$,  \protect\mydiamond[none]\,: $\delta_h^{3}$, \protect\mytriangle[none]\,: $\delta_h^{cos}$, \protect\mysquare[none]\,: $\delta_h^{G}$, \protect\dashedrule\,: first order convergence.}
\label{fig:err_fil_poiss}
\end{figure}
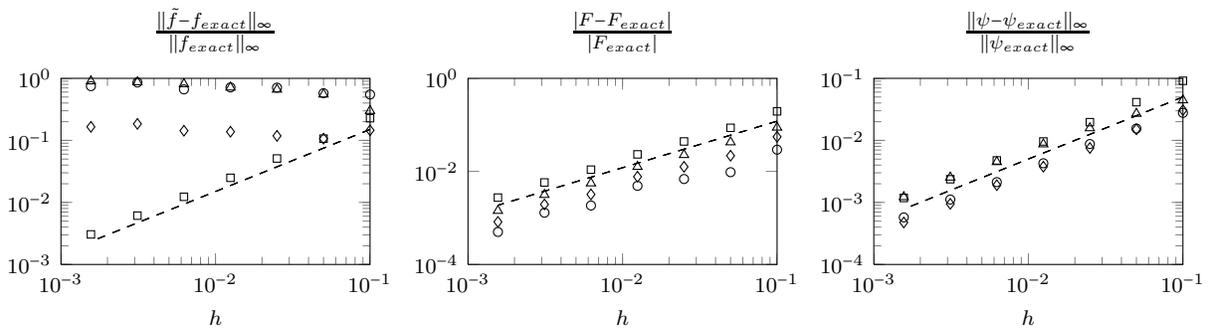

Figure \ref{fig:source_fil_plot} shows the filtered source terms as a function of arc length along the cylinder. By comparison with Figure \ref{fig:source_plot}, it is clear that the filtered surface stresses are better representations of $f_{ex}$ than their unfiltered counterparts. Moreover, note from Figure \ref{fig:source_fil_plot} that the approximation to $f_{ex}$ improves as the smoothness of the smeared delta function increases. This argument is shown quantitatively by the error plot from Figure \ref{fig:err_fil_poiss}. Indeed, the infinitely differentiable $\delta_h^G$ yields an $\tilde{f}$ that converges to $f_{ex}$. The inability to compute convergent source terms using $\delta_h^{hat}$, $\delta_h^3$, and $\delta_h^{cos}$ stems from the slow decay rate of the coefficients $\alpha_j^{ex}$. By contrast, accurate approximations to $f_{ex}$ can be obtained for $\delta_h^G$ by simply removing the high index coefficients of $\alpha_j$.

Note also that it is only the smoothness of the smeared delta functions that matters; $\delta_h^{hat}$, $\delta_h^{3}$, and $\delta_h^G$ all satisfy the same number of discrete moment conditions, and the derivative of $\delta_h^{3}$ satisfies two more discrete moment conditions than $\delta_h^G$. Last, see from Figure \ref{fig:err_fil_poiss} that filtering does not affect $F$ by virtue of the way $\tilde{E}H$ was constructed, and that the error in the solution $\psi$ is unchanged because computing $\tilde{f}$ is a post-processing step. For these reasons, we may write $F$ and $\psi$ without the tilde.

It is worth mentioning other possibilities for accurately computing source terms. First, there might be adequately differentiable functions of narrower support than $\delta_h^G$ that are sufficiently compact in Fourier space to provide convergent source terms. Second, one may use standard regularization techniques that have been developed for first-kind integral equations, such as Tikhonov reguarization, to compute convergent source terms irrespective of delta function. The difficulty in using these techniques is that they involve a free parameter, and our experience has been that a costly SVD is required to determine this parameter so that the source term converges.

\section{Extension to accurately computing surface stresses and forces}

This section shows that surface velocity-based IB methods also require the solution of a discrete integral equation to compute the surface stresses on the immersed body. Therefore, the same conclusion that smoother smeared delta functions lead to faster decay of coefficients for the exact surface stresses holds. Moreover, sufficiently smooth smeared delta functions may be used in combination with the filter $\tilde{E}H$ to obtain surface stresses and forces that converge to the actual stresses and forces on the immersed body.

The nondimensionalized Navier-Stokes equations are considered here on a domain $\Omega$ containing a body whose boundary is denoted by $\Gamma$. The governing equations for surface velocity-based IB methods are written as
\begin{gather}
\frac{\partial{\textbf{u}}}{\partial t} + \textbf{u}\cdot\nabla \textbf{u} = -\nabla p + \frac{1}{Re} \nabla^2\textbf{u} + \int_{\Gamma} \textbf{f}(\boldsymbol\xi(s',t)) \delta(\textbf{x} - \boldsymbol\xi(s',t) ) ds'
\label{eqn:NS1} \\
\nabla \cdot \textbf{u} = 0
\label{eqn:NS2} \\
\int_\Omega \textbf{u}(\textbf{x}) \delta(\textbf{x} - \boldsymbol\xi(s,t)) d\textbf{x} = \textbf{u}^\Gamma(\boldsymbol\xi(s,t), t)
\label{eqn:NS3}
\end{gather}

where $\textbf{f}(\boldsymbol\xi(s',t))$ represents the surface stresses that arise to enforce the boundary condition (\ref{eqn:NS3}). As with the previous section, all IB methods replace the Dirac delta functions in (\ref{eqn:NS1}) and (\ref{eqn:NS3}) with smeared delta functions $\delta_h$.

It is well known that many discretizations of (\ref{eqn:NS1})--(\ref{eqn:NS3}) involve solving a discrete Poisson or Poisson-like equation for either the pressure (primitive variable formulations) or for the streamfunction (vorticity-streamfunction formulations). An analogous situation occurs for the surface stresses, except that the equation is an integral equation. This can be seen by multiplying (\ref{eqn:NS1}) by $\delta_h(\textbf{x}-\boldsymbol\xi(s,t))$ and integrating over the domain. Doing this gives
\begin{equation}
\begin{gathered}
\int_\Omega \int_{\Gamma} \textbf{f}(\boldsymbol\xi(s',t)) \delta_h(\textbf{x} - \boldsymbol\xi(s,t) ) \delta_h(\textbf{x}-\boldsymbol\xi(s',t) )  ds' d\textbf{x} = \\
\int_\Omega \left[ \left( \frac{\partial }{\partial t} - \frac{1}{Re} \nabla^2 \right) \textbf{u}(\textbf{x}) + \textbf{u}\cdot\nabla\textbf{u} + \nabla p \right] \delta_h( \textbf{x}-\boldsymbol\xi(s,t) )d\textbf{x} 
\end{gathered}
\label{eqn:int_eqn}
\end{equation}

The key point is that all IB methods replace the delta function with a smeared delta function in the governing equations. Had the Dirac delta function been kept, the integral equation (\ref{eqn:int_eqn}) would trivially reduce to an expression for the surface stresses $\textbf{f}(\boldsymbol\xi(s,t))$. As in (\ref{eqn:source_sol}), the integral operator of (\ref{eqn:int_eqn}) has an unbounded inverse because it contains a continuous kernel for any finite $h$. 

Many discretizations of (\ref{eqn:NS1})--(\ref{eqn:NS3}) involve solving a discretized integral equation of the first kind for the surface stresses. Spatially discretizing (\ref{eqn:NS1})--(\ref{eqn:NS3}) leads to a system of differential algebraic equations given by 
\begin{gather}
M \frac{du}{dt} + \mathcal{N}(u) = -Gp + Lu + Hf \label{eqn:disc_1} \\
Du = 0 \label{eqn:disc_2} \\
Eu = u^\Gamma \label{eqn:disc_3}
\end{gather}
where $u,p,$ and $f$ denote the spatially discrete velocity, pressure, and surface stresses; $M$ is the (diagonal) mass matrix; $\mathcal{N}(u)$ is a discretization of the nonlinear term; $G$, $L$, and $D$ are discretizations of the gradient, Laplacian, and divergence operators, respectively; and $H(\cdot)$ and $E(\cdot)$ are discretizations of the operations $\int_\Gamma (\cdot) \delta_h(\textbf{x}-\boldsymbol\xi) ds$ and $\int_\Omega (\cdot) \delta_h(\textbf{x}-\boldsymbol\xi) d\textbf{x}$, respectively.

Consider a time discretization that treats the nonlinear term explicitly and the viscous term implicitly. Then (\ref{eqn:disc_1})--(\ref{eqn:disc_3}) become a linear system of equations of the form
\begin{equation}
\begin{bmatrix} A & G & H \\ D & 0 & 0 \\ E & 0 & 0 \end{bmatrix}  \begin{bmatrix} u_{n+1} \\ p_{n+k_1} \\ f_{n+k_2} \end{bmatrix} = \begin{bmatrix} r_1 \\ r_2 \\ u^\Gamma_{n+1} \end{bmatrix}
\label{eqn:disc_sys}
\end{equation}
where $0< k_1, k_2 \le 1$, $A = \frac{1}{\Delta t} M - \alpha L$ ($\alpha \in \mathbb{R}$) comes from the implicit treatment of the viscous term, and $r_1$ and $r_2$ are known right hand side terms arising from the explicit time discretization and boundary conditions of the spatial derivative operators.

The system (\ref{eqn:disc_sys}) is valid for a variety of discretizations. Multistep methods lead to a system of the form of (\ref{eqn:disc_sys}), and many Runge-Kutta methods involve solving a system such as (\ref{eqn:disc_sys}) at each stage. If the viscous term were treated explicitly then $A$ would be replaced with $\frac{1}{\Delta t} M$, though none of the ensuing conclusions would be affected by this change.

The matrix in (\ref{eqn:disc_sys}) may be factorized to give a set of equations for $p_{n+k_1}$ and $f_{n+k_2}$, after which substitution yields an equation for the surface stresses at the desired time step:
\begin{equation}
EBHf_{n+k_2} = r_3
\label{eqn:disc_int}
\end{equation}
where $r_3$ is known and $B = (A^{-1}G(DA^{-1}G)^{-1}D - I) A^{-1}$. The form of $B$ arises because of the time discretization of the system (\ref{eqn:disc_1})--(\ref{eqn:disc_3}) and the factorization of (\ref{eqn:disc_sys}). Equation (\ref{eqn:disc_int}) is an approximation of the continuous equation (\ref{eqn:int_eqn}), and therefore is a discrete integral equation of the first kind. Thus, the logic of section 2 applies: smoother delta functions may be used to expand the exact surface stresses on the body using very few terms, and may therefore be combined with the filter $\tilde{E}H$ to compute accurate surface stresses and forces.

It should be mentioned that the IB method of Colonius and Taira \cite{Colonius:2008dr} is formulated in a vorticity streamfunction framework. It can be shown that this formulation still leads to a discrete integral equation of the first kind whose kernel is modified from (\ref{eqn:int_eqn}) by the presence of discrete curl operators. The conclusions of section 2 are still applicable despite this difference.

The above derivation of the discrete integral equation for the surface stresses does not apply to all surface velocity-based IB methods. Some methods arrive at the equation for the surface stresses by approximation rather than a formal time discretization of (\ref{eqn:disc_1})--(\ref{eqn:disc_3}) \cite{Uhlmann:2005gf,Huang:2009ic,Zhang:2007ez,Yang:2009fo}. In any case, the equation for the surface stresses used by these methods is a discrete integral equation of the first kind. These methods compute the stresses at the desired time step by evaluating the discrete momentum equations on the immersed surface at a previous time. In the notation of this work, this may be written as
\begin{equation}
EHf_{n+1} = \frac{u^\Gamma_{n+1}-Eu_n}{\Delta t} + E(\mathcal{N}(u_{n+k_3}) + Gp_{n+k_4} + Lu_{n+k_5})
\label{eqn:disc_int2}
\end{equation}
where $0\le k_3,k_4,k_5<1$, and all terms on the right hand side are known. This is a discrete integral equation of the first kind whose kernel corresponds to that of (\ref{eqn:int_eqn}).

As an approximation, references \cite{Uhlmann:2005gf,Huang:2009ic,Zhang:2007ez,Yang:2009fo} replace the matrix $EH$ in (\ref{eqn:disc_int2}) with the identity matrix, which corresponds to replacing the kernel in (\ref{eqn:int_eqn}) with an invertible kernel given by the Dirac delta functions. This approximation produces non-convergent surface stresses and forces, though Yang \emph{et al.}\ \cite{Yang:2009fo} reduced the error in the surface forces. The methods of section 2 may be used to obtain convergent surface stresses and forces from the discrete integral equation (\ref{eqn:disc_int2}).

In the ensuing part of this work, we use the IBPM \cite{Colonius:2008dr} to illustrate that computing surface stresses and forces using the filter $\tilde{E}H$ leads to increasingly accurate surface stresses as the smoothness of the smeared delta function is increased. We further show that a sufficiently smooth smeared delta function may be used to obtain convergent stresses and forces. These results are demonstrated for multiple test problems.

\section{An impulsively rotated cylinder}

Consider an infinitely long (2-D), infinitely thin cylinder of radius $R$ in a quiescent fluid that is impulsively brought from rest to constant angular velocity $\omega$. Fluid exists inside and outside of the cylinder. All quantities in this section are dimensionless: length scales are nondimensionalized by $R$, velocities are nondimensionalized by $\omega R$, and time is nondimensionalized by $\omega$. 

The exact velocity field is in the azimuthal direction, and is given in polar coordinates by $\textbf{u}_{ex} = u_{ex}(r,t) \textbf{e}_\theta$. It may be written as
\begin{equation}
u_{ex} (r,t) = 
	\begin{cases} 
		r + 2\sum_{n=1}^\infty \frac{J_1\left(\sqrt{\lambda_n} r \right) }{\sqrt{\lambda_n} J_0(\sqrt{\lambda_n} ) } e^{ \frac{-\lambda_n t}{ Re} } ,\qquad r \le 1 \vspace*{2mm} \\ 
		\mathcal{L}^{-1} \left[ \frac{K_1\left(r \sqrt{s Re} \right) }{s K_1(\sqrt{s Re} )}  \right], \qquad r > 1
	\end{cases} 
	\label{eqn:ex_vel}
\end{equation}

In the above, $Re$ is the Reynolds number; $J_p$ is the $p^{th}$ Bessel function of the first kind; $\sqrt{\lambda_n}$ is the $n^{th}$ root of $J_1$; $K_1$ is the $1^{st}$ modified Bessel function of the first kind; and $\mathcal{L}^{-1}[\cdot]$ represents the inverse Laplace transform with respect to the variable $s$. 

The exact surface stress is also in the azimuthal direction ($\textbf{f}_{ex}  = f _{ex} \textbf{e}_\theta$), and is given by summing the contributions on the inside and outside of the cylinder surface:
\begin{align}
f_{ex}(t) &= \frac{2}{Re} \left[ \frac{\partial }{\partial r}\left( \frac{u_{ex}}{r} \right) \right]_{r = 1^-} + \frac{2}{Re} \left[ \frac{\partial }{\partial r}\left( \frac{u_{ex}}{r} \right) \right]_{r = 1^+} \label{eqn:ex_force_pre}\\
& = \frac{4}{Re} \sum_{n=1}^\infty  e^{ \frac{-\lambda_n t}{ Re} } + \frac{2}{Re} \left[ \frac{\partial }{\partial r}\left( \frac{u_{ex}}{r} \right) \right]_{r = 1^+}
\label{eqn:ex_force}
\end{align}
The second term on the right hand side of (\ref{eqn:ex_force}) is difficult to express analytically by virtue of the inverse Laplace transform in (\ref{eqn:ex_vel}), but it can be evaluated using standard numerical routines. Note that the exact stresses are not spatially constant in the Cartesian coordinate system in which the IBPM is formulated, which makes this model problem a more stringent test than if the numerical solution was obtained using a cylindrical polar coordinate system.

The exact surface force in the azimuthal direction ($F_{ex} $) is obtained by integrating (\ref{eqn:ex_force}) along the surface of the cylinder:
\begin{equation}
F_{ex} (t) = 2\pi f_{ex} (t) 
\label{eqn:net_force}
\end{equation}

The quantities in (\ref{eqn:ex_vel}), (\ref{eqn:ex_force}), and (\ref{eqn:net_force}) were evaluated using standard MATLAB routines, and all quantities were converged to within $10^{-10}$. We compare this exact solution to the IBPM using the smeared delta functions introduced in section 2. We ran tests for Reynolds numbers ranging from $Re = 10$ to $Re = 200$. In the interest of brevity, we primarily show results for $Re = 10$, with supplementary results given for $Re = 200$. 

All simulations used a multidomain approach: fine grids were placed near the immersed body and coarser grids were employed as distance from the immersed body increased. In all results shown below, the cylinder of dimensionless radius $1$ was centered at $[0,0]$; the finest mesh was placed on a subdomain of size $[-2.5,2.5]\times[-2.5,2.5]$, and the total flow domain size was $[-20,20]\times[-20,20]$. The grid spacing on the immersed surface was selected to match that of the $[-2.5,2.5]\times[-2.5,2.5]$ sub-domain, and the time step was selected so that the CFL number with respect to the angular velocity of the cylinder was kept at 0.1. In what follows, $h$ is defined as the grid spacing on the $[-2.5,2.5]\times[-2.5,2.5]$ subdomain.

\begin{figure}[h!]
\setlength{\figH}{0.35\textwidth}
\setlength{\figW}{0.93\textwidth}
\centering
     \input{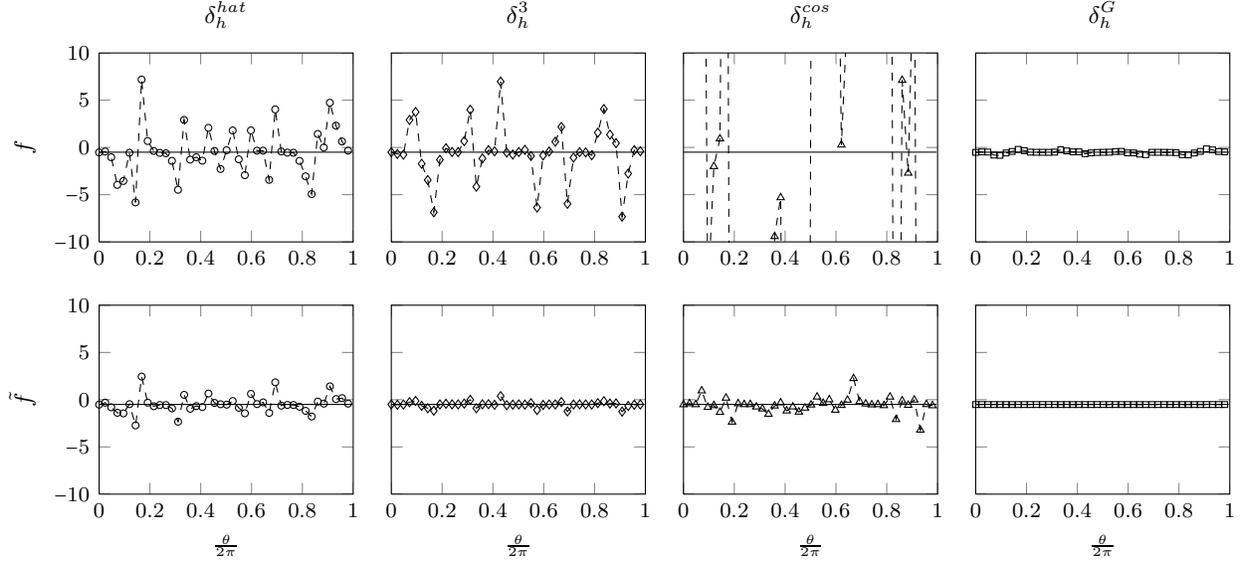}
    \caption[waste]{Top row: tangential surface stress without filtering ($f$) versus arc length along the cylinder for the rotating cylinder problem at $Re = 10$. Bottom row: filtered surface stresses ($\tilde{f}$) versus arc length along the cylinder at $Re = 10$; \protect\solidrule: $f_{ex}$. All plots used $h = 5/200$.}
    \label{fig:f_rot_cyl}
\end{figure}

Figure \ref{fig:f_rot_cyl} demonstrates that for $Re = 10$, the filtered stresses are better approximations to $f_{ex}$ than the unfiltered stresses. Morever, the quality of the approximation of the filtered stress is better for smoother smeared delta functions (see also the error in the filtered stresses from Figure \ref{fig:err_rot_cyl_Re10}). Indeed, the use of $\delta_h^G$ leads to filtered surface stresses that converge to the analytical solution $f_{ex}$.  

\begin{figure}[h!]
\setlength{\figH}{0.15\textwidth}
\setlength{\figW}{0.98\textwidth}
\centering
%
%
\begin{tikzpicture}

\begin{axis}[%
width=0.262\figW,
height=\figH,
at={(0\figW,0\figH)},
scale only axis,
xmode=log,
xmin=0.001,
xmax=0.1,
xminorticks=true,
xlabel={$h$},
ymode=log,
ymin=0.001,
ymax=1000,
yminorticks=true,
axis background/.style={fill=white},
title style={font=\bfseries},
title={$\frac{||\tilde{f} - f_{exact}||_\infty}{||f_{exact}||_\infty}$},
ticklabel style={font=\footnotesize},ylabel style={font=\footnotesize},xlabel style={font=\footnotesize},title style={font=\normalsize}
]
\addplot [color=black,mark size=1.7pt,only marks,mark=*,mark options={solid,fill=white},forget plot]
  table[row sep=crcr]{%
0.0208333333333333	2.18660566333979\\
0.01	7.18201228391791\\
0.005	7.57162805374547\\
0.0025	11.1440441188571\\
0.00125	22.021786381961\\
};
\addplot [color=black,mark size=2pt,only marks,mark=diamond*,mark options={solid,fill=white},forget plot]
  table[row sep=crcr]{%
0.0208333333333333	1.2557315216727\\
0.01	1.25498859504959\\
0.005	2.19699085004663\\
0.0025	3.41362011220428\\
0.00125	3.13000926822014\\
};
\addplot [color=black,mark size=2pt,only marks,mark=triangle*,mark options={solid,fill=white},forget plot]
  table[row sep=crcr]{%
0.0208333333333333	1.36355596697485\\
0.01	7.89889968569564\\
0.005	2.88535910691101\\
0.0025	9.09099269156747\\
0.00125	10.3156580010667\\
};
\addplot [color=black,mark size=1.4pt,only marks,mark=square*,mark options={solid,fill=white},forget plot]
  table[row sep=crcr]{%
0.0208333333333333	0.621319000435923\\
0.01	0.0361296074884602\\
0.005	0.0135500887747912\\
0.0025	0.00706043732222061\\
0.00125	0.00328648395211933\\
};
\addplot [color=black,dashed,forget plot]
  table[row sep=crcr]{%
0.0208333333333333	0.09375\\
0.01	0.045\\
0.005	0.0225\\
0.0025	0.01125\\
0.00125	0.005625\\
};
\end{axis}

\begin{axis}[%
width=0.262\figW,
height=\figH,
at={(0.345\figW,0\figH)},
scale only axis,
xmode=log,
xmin=0.001,
xmax=0.1,
xminorticks=true,
xlabel={$h$},
ymode=log,
ymin=0.0001,
ymax=0.1,
ytick={0.0001, 0.001,  0.01,   0.1},
yminorticks=true,
axis background/.style={fill=white},
title={$\frac{|F - F_{exact}|}{|F_{exact}|}$},
ticklabel style={font=\footnotesize},ylabel style={font=\footnotesize},xlabel style={font=\footnotesize},title style={font=\normalsize}
]
\addplot [color=black,mark size=1.7pt,only marks,mark=*,mark options={solid,fill=white},forget plot]
  table[row sep=crcr]{%
0.0208333333333333	0.0140532169341069\\
0.01	0.00137952833075932\\
0.005	0.00166780941616451\\
0.0025	0.00130821411173737\\
0.00125	0.000684881515781044\\
};
\addplot [color=black,mark size=2pt,only marks,mark=diamond*,mark options={solid,fill=white},forget plot]
  table[row sep=crcr]{%
0.0208333333333333	0.00911673206137508\\
0.01	0.000843288579148276\\
0.005	0.001865172884025\\
0.0025	0.00127225675914402\\
0.00125	0.000704230145007111\\
};
\addplot [color=black,mark size=2pt,only marks,mark=triangle*,mark options={solid,fill=white},forget plot]
  table[row sep=crcr]{%
0.0208333333333333	0.00286058592873571\\
0.01	0.0073903254189183\\
0.005	0.00370852402368309\\
0.0025	0.00337214871530234\\
0.00125	0.00161037540049262\\
};
\addplot [color=black,mark size=1.4pt,only marks,mark=square*,mark options={solid,fill=white},forget plot]
  table[row sep=crcr]{%
0.0208333333333333	0.0458374363283334\\
0.01	0.00397959774257249\\
0.005	0.00201371716015492\\
0.0025	0.00196281050674316\\
0.00125	0.00121772497777864\\
};
\addplot [color=black,dashed,forget plot]
  table[row sep=crcr]{%
0.0208333333333333	0.0104166666666667\\
0.01	0.005\\
0.005	0.0025\\
0.0025	0.00125\\
0.00125	0.000625\\
};
\end{axis}

\begin{axis}[%
width=0.262\figW,
height=\figH,
at={(0.689\figW,0\figH)},
scale only axis,
xmode=log,
xmin=0.001,
xmax=0.1,
xminorticks=true,
xlabel={$h$},
ymode=log,
ymin=0.001,
ymax=1,
yminorticks=true,
axis background/.style={fill=white},
title style={font=\bfseries},
title={$\frac{||u - u_{exact}||_\infty}{||u_{exact}||_\infty}$},
ticklabel style={font=\footnotesize},ylabel style={font=\footnotesize},xlabel style={font=\footnotesize},title style={font=\normalsize}
]
\addplot [color=black,mark size=1.7pt,only marks,mark=*,mark options={solid,fill=white},forget plot]
  table[row sep=crcr]{%
0.0208333333333333	0.163237542263635\\
0.01	0.102590074250526\\
0.005	0.0482347229847793\\
0.0025	0.025759067913755\\
0.00125	0.0149043009088213\\
};
\addplot [color=black,mark size=2pt,only marks,mark=diamond*,mark options={solid,fill=white},forget plot]
  table[row sep=crcr]{%
0.0208333333333333	0.239699831625197\\
0.01	0.103062624909607\\
0.005	0.0520199362160785\\
0.0025	0.0266429729586539\\
0.00125	0.0137686654292203\\
};
\addplot [color=black,mark size=2pt,only marks,mark=triangle*,mark options={solid,fill=white},forget plot]
  table[row sep=crcr]{%
0.0208333333333333	0.315567773772305\\
0.01	0.170024247610389\\
0.005	0.0897106769735696\\
0.0025	0.0565686242363931\\
0.00125	0.0319416436499391\\
};
\addplot [color=black,mark size=1.4pt,only marks,mark=square*,mark options={solid,fill=white},forget plot]
  table[row sep=crcr]{%
0.0208333333333333	0.438773484497303\\
0.01	0.19826166023585\\
0.005	0.101727594941829\\
0.0025	0.0514799917389073\\
0.00125	0.0257356795755836\\
};
\addplot [color=black,dashed,forget plot]
  table[row sep=crcr]{%
0.0208333333333333	0.260416666666667\\
0.01	0.125\\
0.005	0.0625\\
0.0025	0.03125\\
0.00125	0.015625\\
};
\end{axis}
\end{tikzpicture}%
    \caption[waste]{Errors in $\tilde{f}$, $F$, and $u$ versus grid spacing ($h$) for the rotating cylinder problem at $Re = 200$. \protect\mycircle[none]\,: $\delta_h^{hat}$,  \protect\mydiamond[none]\,: $\delta_h^{3}$, \protect\mytriangle[none]\,: $\delta_h^{cos}$, \protect\mysquare[none]\,: $\delta_h^{G}$, \protect\dashedrule\,: first order convergence.}
    \label{fig:err_rot_cyl_Re200}
\end{figure}
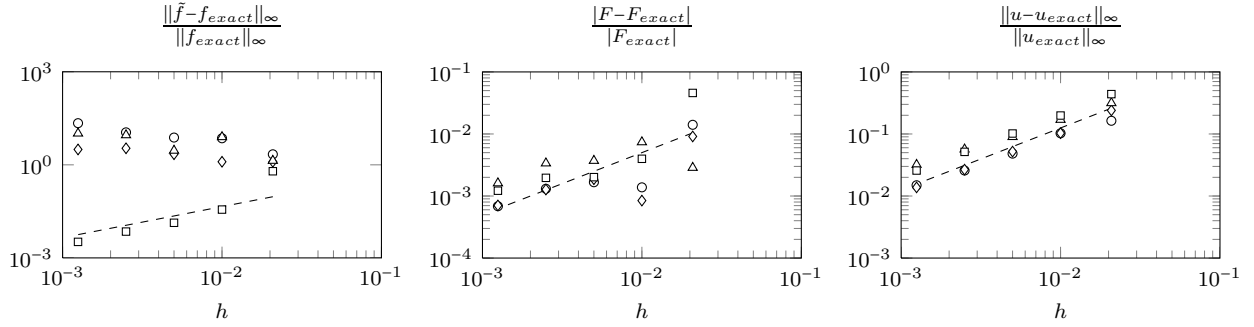

In analogy with section 2, the surface forces converge irrespective of smeared delta function (see Figures \ref{fig:err_rot_cyl_Re10} \ref{fig:F_rot_cyl}). This is a consequence of solving the discrete integral equation derived in section 3 to explicitly enforce the boundary condition. The surface velocity-based IB methods that approximately enforce this condition are known to obtain non-convergent surface forces and stresses \cite{Yang:2009fo}. Note also that the velocity field converges at first order for all smeared delta functions. In keeping with the notation of section 2, tildes are not placed on the variables $F$ and $u$ to emphasize that these quantities are not affected by the filtering procedure. Figure \ref{fig:err_rot_cyl_Re200} shows the errors in $\tilde{f}$, $F$, and $u$ at $Re = 200$ to highlight the applicability of these results over a range of Reynolds numbers.

\begin{figure}[h!]
\setlength{\figH}{0.15\textwidth}
\setlength{\figW}{0.98\textwidth}
\centering
%
%
\begin{tikzpicture}

\begin{axis}[%
width=0.262\figW,
height=\figH,
at={(0\figW,0\figH)},
scale only axis,
xmode=log,
xmin=0.001,
xmax=0.1,
xminorticks=true,
xlabel={$h$},
ymode=log,
ymin=0.001,
ymax=1000,
yminorticks=true,
axis background/.style={fill=white},
title style={font=\bfseries},
title={$\frac{||\tilde{f} - f_{exact}||_\infty}{||f_{exact}||_\infty}$},
ticklabel style={font=\footnotesize},ylabel style={font=\footnotesize},xlabel style={font=\footnotesize},title style={font=\normalsize}
]
\addplot [color=black,mark size=1.7pt,only marks,mark=*,mark options={solid,fill=white},forget plot]
  table[row sep=crcr]{%
0.0208333333333333	0.924001278887112\\
0.01	5.38158543919383\\
0.005	8.27794575353563\\
0.0025	11.6622090820703\\
0.00125	22.4667887730173\\
};
\addplot [color=black,mark size=2.0pt,only marks,mark=diamond*,mark options={solid,fill=white},forget plot]
  table[row sep=crcr]{%
0.0208333333333333	0.184947692183415\\
0.01	1.2504388131035\\
0.005	2.29767905024481\\
0.0025	3.45070999465176\\
0.00125	3.17386799815074\\
};
\addplot [color=black,mark size=2.0pt,only marks,mark=triangle*,mark options={solid,fill=white},forget plot]
  table[row sep=crcr]{%
0.0208333333333333	1.92528618388396\\
0.01	9.36858236018287\\
0.005	7.01235377547013\\
0.0025	15.7954994801151\\
0.00125	16.8554217335565\\
};
\addplot [color=black,mark size=1.4pt,only marks,mark=square*,mark options={solid,fill=white},forget plot]
  table[row sep=crcr]{%
0.0208333333333333	0.141851412524029\\
0.01	0.0681104828342351\\
0.005	0.0325364941963821\\
0.0025	0.0169301915551745\\
0.00125	0.00828267313736119\\
};
\addplot [color=black,dashed,forget plot]
  table[row sep=crcr]{%
0.0208333333333333	0.09375\\
0.01	0.045\\
0.005	0.0225\\
0.0025	0.01125\\
0.00125	0.005625\\
};
\end{axis}

\begin{axis}[%
width=0.262\figW,
height=\figH,
at={(0.345\figW,0\figH)},
scale only axis,
xmode=log,
xmin=0.001,
xmax=0.1,
xminorticks=true,
xlabel={$h$},
ymode=log,
ymin=0.001,
ymax=1,
ytick={0.001,  0.01,   0.1,     1},
yminorticks=true,
axis background/.style={fill=white},
title={$\frac{|F - F_{exact}|}{|F_{exact}|}$},
ticklabel style={font=\footnotesize},ylabel style={font=\footnotesize},xlabel style={font=\footnotesize},title style={font=\normalsize}
]
\addplot [color=black,mark size=1.7pt,only marks,mark=*,mark options={solid,fill=white},forget plot]
  table[row sep=crcr]{%
0.0208333333333333	0.0392169305038114\\
0.01	0.0250998065991953\\
0.005	0.0119904097238012\\
0.0025	0.00628441352676086\\
0.00125	0.00311161563541865\\
};
\addplot [color=black,mark size=2pt,only marks,mark=diamond*,mark options={solid,fill=white},forget plot]
  table[row sep=crcr]{%
0.0208333333333333	0.062846993247465\\
0.01	0.0294246267176827\\
0.005	0.0147266189795009\\
0.0025	0.00738223411991213\\
0.00125	0.00367946462223694\\
};
\addplot [color=black,mark size=2pt,only marks,mark=triangle*,mark options={solid,fill=white},forget plot]
  table[row sep=crcr]{%
0.0208333333333333	0.104475567708722\\
0.01	0.0488950644228658\\
0.005	0.0237095387656401\\
0.0025	0.0127403347666986\\
0.00125	0.00616704051931721\\
};
\addplot [color=black,mark size=1.4pt,only marks,mark=square*,mark options={solid,fill=white},forget plot]
  table[row sep=crcr]{%
0.0208333333333333	0.12471335880942\\
0.01	0.0582990015970903\\
0.005	0.0287923890625427\\
0.0025	0.0143079740826766\\
0.00125	0.00713209518281323\\
};
\addplot [color=black,dashed,forget plot]
  table[row sep=crcr]{%
0.0208333333333333	0.0729166666666667\\
0.01	0.035\\
0.005	0.0175\\
0.0025	0.00875\\
0.00125	0.004375\\
};
\end{axis}

\begin{axis}[%
width=0.262\figW,
height=\figH,
at={(0.689\figW,0\figH)},
scale only axis,
xmode=log,
xmin=0.001,
xmax=0.1,
xminorticks=true,
xlabel={$h$},
ymode=log,
ymin=0.001,
ymax=1,
yminorticks=true,
axis background/.style={fill=white},
title style={font=\bfseries},
title={$\frac{||u - u_{exact}||_\infty}{||u_{exact}||_\infty}$},
ticklabel style={font=\footnotesize},ylabel style={font=\footnotesize},xlabel style={font=\footnotesize},title style={font=\normalsize}
]
\addplot [color=black,mark size=1.7pt,only marks,mark=*,mark options={solid,fill=white},forget plot]
  table[row sep=crcr]{%
0.0208333333333333	0.0505125860861132\\
0.01	0.0332863678062193\\
0.005	0.0164513966901594\\
0.0025	0.00881354812021773\\
0.00125	0.00501595219539963\\
};
\addplot [color=black,mark size=2pt,only marks,mark=diamond*,mark options={solid,fill=white},forget plot]
  table[row sep=crcr]{%
0.0208333333333333	0.0720533004169963\\
0.01	0.0326499262634257\\
0.005	0.0169503567729967\\
0.0025	0.00872144419411891\\
0.00125	0.00459119599188873\\
};
\addplot [color=black,mark size=2pt,only marks,mark=triangle*,mark options={solid,fill=white},forget plot]
  table[row sep=crcr]{%
0.0208333333333333	0.142871852237979\\
0.01	0.0812433187464091\\
0.005	0.0435863418074083\\
0.0025	0.0289549387710514\\
0.00125	0.0161159840129297\\
};
\addplot [color=black,mark size=1.4pt,only marks,mark=square*,mark options={solid,fill=white},forget plot]
  table[row sep=crcr]{%
0.0208333333333333	0.136559295080995\\
0.01	0.0646311636533623\\
0.005	0.0329141727259996\\
0.0025	0.0167117583355334\\
0.00125	0.00842330385457202\\
};
\addplot [color=black,dashed,forget plot]
  table[row sep=crcr]{%
0.0208333333333333	0.09375\\
0.01	0.045\\
0.005	0.0225\\
0.0025	0.01125\\
0.00125	0.005625\\
};
\end{axis}
\end{tikzpicture}%
    \caption[waste]{Errors in $\tilde{f}$, $F$, and $u$ versus grid spacing ($h$) for the rotating cylinder problem at $Re = 10$. \protect\mycircle[none]\,: $\delta_h^{hat}$,  \protect\mydiamond[none]\,: $\delta_h^{3}$, \protect\mytriangle[none]\,: $\delta_h^{cos}$, \protect\mysquare[none]\,: $\delta_h^{G}$, \protect\dashedrule\,: first order convergence.}
    \label{fig:err_rot_cyl_Re10}
\end{figure}
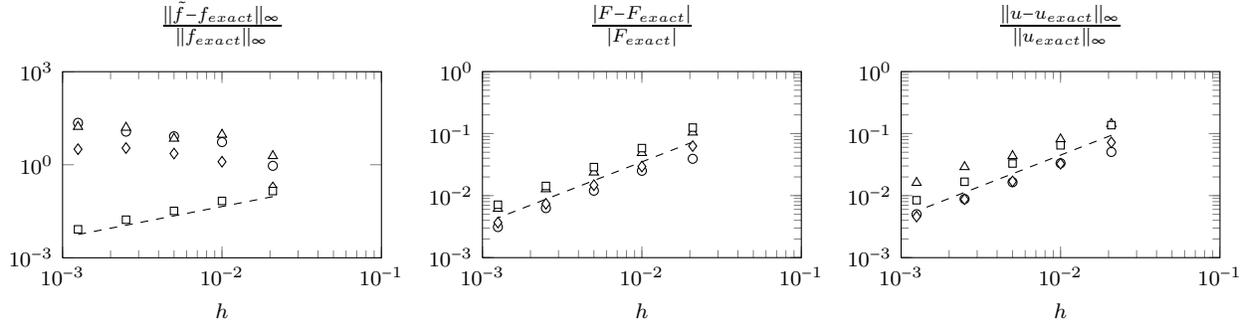

\begin{figure}[h!]
\setlength{\figH}{0.15\textwidth}
\setlength{\figW}{0.98\textwidth}
\centering
%
%
\begin{tikzpicture}

\begin{axis}[%
width=0.22\figW,
height=\figH,
at={(0\figW,0\figH)},
scale only axis,
xmin=0,
xmax=5,
xtick={0, 1, 2, 3, 4, 5},
xlabel={$t$},
ymin=-7,
ymax=0,
ytick={-6, -4, -2,  0},
ylabel={$F$},
axis background/.style={fill=white},
title style={font=\bfseries},
title={$\delta_h^{hat}$},
ticklabel style={font=\footnotesize},ylabel style={font=\small},xlabel style={font=\small},title style={font=\footnotesize}
]
\addplot [color=black,mark size=1.3pt,only marks,mark=o,mark options={solid},forget plot]
  table[row sep=crcr]{%
0.002	-317.907254229243\\
0.128	-12.7564780876759\\
0.254	-9.12086903662917\\
0.38	-7.52652773746274\\
0.506	-6.58734475606873\\
0.632	-5.95487962213713\\
0.758	-5.49489543434046\\
0.884	-5.14332910618454\\
1.01	-4.86528181196144\\
1.136	-4.63991115562139\\
1.262	-4.45387126826161\\
1.388	-4.2981258233256\\
1.514	-4.16627065131372\\
1.64	-4.05359540219067\\
1.766	-3.95653126202475\\
1.892	-3.87231061559051\\
2.018	-3.79874805648791\\
2.144	-3.7340936280902\\
2.27	-3.67693072990706\\
2.396	-3.62610271066504\\
2.522	-3.58065858018798\\
2.648	-3.53981191526706\\
2.774	-3.50290916208315\\
2.9	-3.46940481648422\\
3.026	-3.43884175625231\\
3.152	-3.41083550695978\\
3.278	-3.38506155829977\\
3.404	-3.36124507598251\\
3.53	-3.33915251400776\\
3.656	-3.31858474687062\\
3.782	-3.29937142556993\\
3.908	-3.2813663244989\\
4.034	-3.26444349439428\\
4.16	-3.24849407357567\\
4.286	-3.23342363852712\\
4.412	-3.21914999755641\\
4.538	-3.20560134929631\\
4.664	-3.19271474228612\\
};
\addplot [color=black,solid,line width=1.0pt,forget plot]
  table[row sep=crcr]{%
0.01	-44.873579537289\\
0.26	-8.96937788456697\\
0.51	-6.53068636643629\\
0.76	-5.45904882619434\\
1.01	-4.83627250237924\\
1.26	-4.42696394600009\\
1.51	-4.13940343963167\\
1.76	-3.92877698477513\\
2.01	-3.76976462432923\\
2.26	-3.64668291497375\\
2.51	-3.54926398859729\\
2.76	-3.470541563757\\
3.01	-3.40569268888518\\
3.26	-3.35130587819702\\
3.51	-3.30494529588915\\
3.76	-3.26484115859708\\
4.01	-3.22969330492944\\
4.26	-3.19853450397446\\
4.51	-3.17063413162055\\
4.76	-3.14543783822262\\
};
\end{axis}

\begin{axis}[%
width=0.22\figW,
height=\figH,
at={(0.24\figW,0\figH)},
scale only axis,
xmin=0,
xmax=5,
xtick={0, 1, 2, 3, 4, 5},
xlabel={$t$},
ymin=-7,
ymax=0,
ytick={-6,-4,-2,0},
yticklabels={\empty},
axis background/.style={fill=white},
title style={font=\bfseries},
title={$\delta_h^{3}$},
ticklabel style={font=\footnotesize},ylabel style={font=\small},xlabel style={font=\small},title style={font=\footnotesize}
]
\addplot [color=black,mark size=1.3pt,only marks,mark=diamond,mark options={solid},forget plot]
  table[row sep=crcr]{%
0.002	-361.821189812848\\
0.128	-12.7515712617786\\
0.254	-9.12264446918183\\
0.38	-7.53004792494996\\
0.506	-6.59166244948234\\
0.632	-5.95967881140165\\
0.758	-5.50004273476948\\
0.884	-5.14876294086709\\
1.01	-4.87097407186503\\
1.136	-4.64584982646974\\
1.262	-4.46005098965733\\
1.388	-4.30454254333505\\
1.514	-4.17291885925115\\
1.64	-4.06046702480391\\
1.766	-3.96361555430794\\
1.892	-3.87959456972166\\
2.018	-3.80621702007181\\
2.144	-3.74173195264164\\
2.27	-3.6847223584369\\
2.396	-3.63403166627517\\
2.522	-3.58870934652844\\
2.648	-3.54796971423515\\
2.774	-3.51116014087292\\
2.9	-3.47773615880955\\
3.026	-3.44724173304005\\
3.152	-3.4192934809701\\
3.278	-3.39356795547067\\
3.404	-3.36979133425742\\
3.53	-3.34773101834067\\
3.656	-3.32718875709869\\
3.782	-3.30799500203904\\
3.908	-3.29000425473239\\
4.034	-3.27309122276065\\
4.16	-3.25714763492086\\
4.286	-3.24207959611613\\
4.412	-3.22780538535667\\
4.538	-3.21425361856869\\
4.664	-3.20136171246716\\
};
\addplot [color=black,solid,line width=1.0pt,forget plot]
  table[row sep=crcr]{%
0.01	-44.873579537289\\
0.26	-8.96937788456697\\
0.51	-6.53068636643629\\
0.76	-5.45904882619434\\
1.01	-4.83627250237924\\
1.26	-4.42696394600009\\
1.51	-4.13940343963167\\
1.76	-3.92877698477513\\
2.01	-3.76976462432923\\
2.26	-3.64668291497375\\
2.51	-3.54926398859729\\
2.76	-3.470541563757\\
3.01	-3.40569268888518\\
3.26	-3.35130587819702\\
3.51	-3.30494529588915\\
3.76	-3.26484115859708\\
4.01	-3.22969330492944\\
4.26	-3.19853450397446\\
4.51	-3.17063413162055\\
4.76	-3.14543783822262\\
};
\end{axis}

\begin{axis}[%
width=0.22\figW,
height=\figH,
at={(0.48\figW,0\figH)},
scale only axis,
xmin=0,
xmax=5,
xtick={0, 1, 2, 3, 4, 5},
xlabel={$t$},
ymin=-7,
ymax=0,
ytick={-6,-4,-2,0},
yticklabels={\empty},
axis background/.style={fill=white},
title style={font=\bfseries},
title={$\delta_h^{cos}$},
ticklabel style={font=\footnotesize},ylabel style={font=\small},xlabel style={font=\small},title style={font=\footnotesize}
]
\addplot [color=black,mark size=1.6pt,only marks,mark=triangle,mark options={solid},forget plot]
  table[row sep=crcr]{%
0.002	-458.293541701963\\
0.128	-12.7591944343271\\
0.254	-9.14144848283292\\
0.38	-7.55247597329762\\
0.506	-6.61598852069915\\
0.632	-5.98522434163813\\
0.758	-5.52647215828455\\
0.884	-5.17589074788509\\
1.01	-4.8986904062711\\
1.136	-4.67408359772603\\
1.262	-4.48874905657001\\
1.388	-4.33365805536631\\
1.514	-4.20240547716405\\
1.64	-4.09027681628615\\
1.766	-3.99369896657004\\
1.892	-3.90990139147732\\
2.018	-3.83669766005693\\
2.144	-3.77233866227585\\
2.27	-3.71541022328816\\
2.396	-3.66475930989311\\
2.522	-3.61943935590277\\
2.648	-3.57866882775093\\
2.774	-3.54179925041859\\
2.9	-3.5082901765542\\
3.026	-3.47768936639563\\
3.152	-3.44961695039068\\
3.278	-3.42375268136704\\
3.404	-3.39982561238348\\
3.53	-3.3776056975782\\
3.656	-3.356896929319\\
3.782	-3.33753171032957\\
3.908	-3.31936622359141\\
4.034	-3.30227661190624\\
4.16	-3.28615581706665\\
4.286	-3.27091095827898\\
4.412	-3.2564611528383\\
4.538	-3.24273570042004\\
4.664	-3.2296725670025\\
};
\addplot [color=black,solid,line width=1.0pt,forget plot]
  table[row sep=crcr]{%
0.01	-44.873579537289\\
0.26	-8.96937788456697\\
0.51	-6.53068636643629\\
0.76	-5.45904882619434\\
1.01	-4.83627250237924\\
1.26	-4.42696394600009\\
1.51	-4.13940343963167\\
1.76	-3.92877698477513\\
2.01	-3.76976462432923\\
2.26	-3.64668291497375\\
2.51	-3.54926398859729\\
2.76	-3.470541563757\\
3.01	-3.40569268888518\\
3.26	-3.35130587819702\\
3.51	-3.30494529588915\\
3.76	-3.26484115859708\\
4.01	-3.22969330492944\\
4.26	-3.19853450397446\\
4.51	-3.17063413162055\\
4.76	-3.14543783822262\\
};
\end{axis}

\begin{axis}[%
width=0.22\figW,
height=\figH,
at={(0.72\figW,0\figH)},
scale only axis,
xmin=0,
xmax=5,
xtick={0, 1, 2, 3, 4, 5},
xlabel={$t$},
ymin=-7,
ymax=0,
ytick={-6,-4,-2,0},
yticklabels={\empty},
axis background/.style={fill=white},
title style={font=\bfseries},
title={$\delta_h^{G}$},
ticklabel style={font=\footnotesize},ylabel style={font=\small},xlabel style={font=\small},title style={font=\small}
]
\addplot [color=black,mark size=1.1pt,only marks,mark=square,mark options={solid},forget plot]
  table[row sep=crcr]{%
0.002	-593.527676263536\\
0.128	-12.7158918477572\\
0.254	-9.127375265976\\
0.38	-7.54531724083676\\
0.506	-6.61173398069492\\
0.632	-5.98263080794646\\
0.758	-5.52506019740146\\
0.884	-5.17546539586369\\
1.01	-4.89918188889418\\
1.136	-4.67547902352344\\
1.262	-4.49105731286307\\
1.388	-4.33689066370617\\
1.514	-4.20656680942981\\
1.64	-4.09536012137204\\
1.766	-3.99968585186319\\
1.892	-3.91676314713529\\
2.018	-3.844397380211\\
2.144	-3.7808335394817\\
2.27	-3.72465368327815\\
2.396	-3.67470283715229\\
2.522	-3.63003396664111\\
2.648	-3.58986620259896\\
2.774	-3.55355256513054\\
2.9	-3.52055467854325\\
3.026	-3.4904227458148\\
3.152	-3.46277955081548\\
3.278	-3.43730758917648\\
3.404	-3.41373865710136\\
3.53	-3.39184538881791\\
3.656	-3.37143435023934\\
3.782	-3.35234038290633\\
3.908	-3.33442195742256\\
4.034	-3.31755734540899\\
4.16	-3.30164145752633\\
4.286	-3.28658322524607\\
4.412	-3.27230342776154\\
4.538	-3.25873288425512\\
4.664	-3.24581094674873\\
};
\addplot [color=black,solid,line width=1.0pt,forget plot]
  table[row sep=crcr]{%
0.01	-44.873579537289\\
0.26	-8.96937788456697\\
0.51	-6.53068636643629\\
0.76	-5.45904882619434\\
1.01	-4.83627250237924\\
1.26	-4.42696394600009\\
1.51	-4.13940343963167\\
1.76	-3.92877698477513\\
2.01	-3.76976462432923\\
2.26	-3.64668291497375\\
2.51	-3.54926398859729\\
2.76	-3.470541563757\\
3.01	-3.40569268888518\\
3.26	-3.35130587819702\\
3.51	-3.30494529588915\\
3.76	-3.26484115859708\\
4.01	-3.22969330492944\\
4.26	-3.19853450397446\\
4.51	-3.17063413162055\\
4.76	-3.14543783822262\\
};
\end{axis}
\end{tikzpicture}%
    \caption[waste]{Tangential surface force, $F$, versus time for the rotating cylinder problem at $Re = 10$; \protect\solidrule: $F_{ex}$. The same grid spacing as in Figure \ref{fig:f_rot_cyl} was used.}
    \label{fig:F_rot_cyl}
\end{figure}
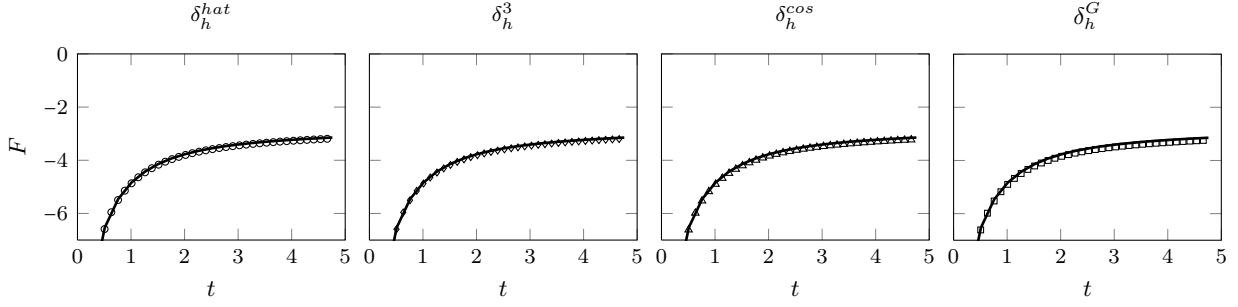

\section{A cylinder in cross-flow}

We now consider the canonical problem of flow over an infinitely long (2D) cylinder of diameter $D$ that is impulsively brought to translation at speed $U$. As with section 4, all quantities are dimensionless; length scales, velocity scales, and time scales are nondimensionalized by $U$, $D$, and $U/D$, respectively. Since there is no known analytical solution to this flow, we will present results at $Re = 200$ to compare with other numerical and experimental results. This flow is well known to exhibit a vortex shedding instability, which we trigger in our simulations using an asymmetric body force at early time. In the interest of brevity, we only present the filtered stresses $\tilde{f}$ for this problem, though the result from sections 2 and 4 that filtering provides better approximations to the physically correct surface stresses remains true here as well. The surface stresses associated with this flow exhibit substantial spatial variation, which attests to the ability of this method to compute convergent surface stresses for a variety of complicated flows.

In all results shown below, the cylinder of dimensionless diameter $1$ was centered at $[0,0]$; the finest mesh was placed on a subdomain of size $[-1.5,2.5]\times[-2,2]$, and the total flow domain size was $[-12,20]\times[-16,16]$. The grid spacing on the immersed surface was selected to match that of the $[-1.5,2.5]\times[-2,2]$ sub-domain, and the time step was selected so that the CFL number with respect to the translational speed of the cylinder was 0.1. In what follows, $h$ is defined as the grid spacing on the $[-1.5,2.5]\times[-2,2]$ subdomain. We define the quantities of interest for this 2-D flow as $\tilde{f} = [\tilde{f}_x, \, \tilde{f}_y]^T$, $F = [C_D, \, C_L]^T$, and $u = [u_x, \, u_y]$, where $C_D$ and $C_L$ denote the dimensionless $x$ and $y$ surface forces, respectively.

\begin{figure}[h!]
\setlength{\figH}{0.3\textwidth}
\setlength{\figW}{0.93\textwidth}
\centering
     \input{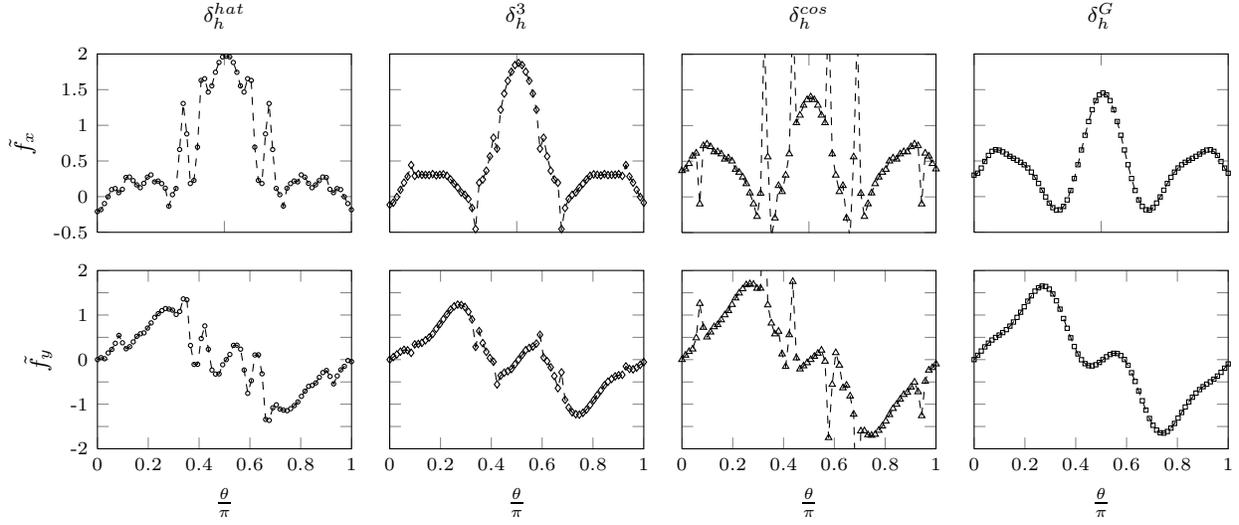}
    \caption[waste]{Filtered $x$ and $y$ component of the surface stress versus arc length along the cylinder for the cylinder in cross-flow problem. All plots used $h = 4/768$.}
    \label{fig:f_trans_cyl}
\end{figure}

\begin{figure}[h!]
\setlength{\figH}{0.15\textwidth}
\setlength{\figW}{0.98\textwidth}
\centering
%
%
\begin{tikzpicture}

\begin{axis}[%
width=0.262\figW,
height=\figH,
at={(0\figW,0\figH)},
scale only axis,
xmode=log,
xmin=0.0001,
xmax=0.01,
xminorticks=true,
xlabel={$h$},
ymode=log,
ymin=0.01,
ymax=10,
ytick={0.01,  0.1,    1,   10},
yminorticks=true,
axis background/.style={fill=white},
title={$\frac{||\tilde{f}_x - \tilde{f}_{x,fine}||_\infty}{||f_{x,fine}||_\infty}$},
ticklabel style={font=\footnotesize},ylabel style={font=\footnotesize},xlabel style={font=\footnotesize},title style={font=\normalsize}
]
\addplot [color=black,mark size=1.4pt,only marks,mark=o,mark options={solid},forget plot]
  table[row sep=crcr]{%
0.00520833333333333	3.04758524456739\\
0.00260416666666667	1.30035751776994\\
0.00130208333333333	0.668418394858334\\
0.000651041666666667	2.17521183272074\\
};
\addplot [color=black,mark size=2.0pt,only marks,mark=diamond,mark options={solid},forget plot]
  table[row sep=crcr]{%
0.00520833333333333	0.422658340831384\\
0.00260416666666667	0.491159728002984\\
0.00130208333333333	0.429730667205591\\
0.000651041666666667	0.564962766903718\\
};
\addplot [color=black,mark size=2.0pt,only marks,mark=triangle,mark options={solid},forget plot]
  table[row sep=crcr]{%
0.00520833333333333	0.433723035458796\\
0.00260416666666667	0.371777515470162\\
0.00130208333333333	2.6979288301609\\
0.000651041666666667	1.74863764680069\\
};
\addplot [color=black,mark size=1.4pt,only marks,mark=square,mark options={solid},forget plot]
  table[row sep=crcr]{%
0.00520833333333333	0.989942236088591\\
0.00260416666666667	0.319711980973698\\
0.00130208333333333	0.1026454003270611\\
0.000651041666666667	0.0515839419849867\\
};
\addplot [color=black,dashed,line width=0.5pt,forget plot]
  table[row sep=crcr]{%
0.00520833333333333	0.364583333333333\\
0.00260416666666667	0.182291666666667\\
0.00130208333333333	0.0911458333333333\\
0.000651041666666667	0.0455729166666667\\
};
\end{axis}

\begin{axis}[%
width=0.262\figW,
height=\figH,
at={(0.345\figW,0\figH)},
scale only axis,
xmode=log,
xmin=0.0001,
xmax=0.01,
xtick={0.0001, 0.001, 0.01},
xminorticks=true,
xlabel={$h$},
ymode=log,
ymin=0.0001,
ymax=0.1,
ytick={0.0001, 0.001, 0.01,  0.1},
yminorticks=true,
axis background/.style={fill=white},
title={$\frac{|C_D - C_{D,fine}|}{|C_{D,fine}|}$},
ticklabel style={font=\footnotesize},ylabel style={font=\footnotesize},xlabel style={font=\footnotesize},title style={font=\normalsize}
]
\addplot [color=black,mark size=1.4pt,only marks,mark=o,mark options={solid},forget plot]
  table[row sep=crcr]{%
0.00520833333333333	0.0108090675697746\\
0.00260416666666667	0.00513769518914113\\
0.00130208333333333	0.00176714519247273\\
0.000651041666666667	0.000682839825802272\\
};
\addplot [color=black,mark size=2.0pt,only marks,mark=diamond,mark options={solid},forget plot]
  table[row sep=crcr]{%
0.00520833333333333	0.00113992670115113\\
0.00260416666666667	0.00212723315968939\\
0.00130208333333333	0.00134150478613551\\
0.000651041666666667	0.000514804190892674\\
};
\addplot [color=black,mark size=2.0pt,only marks,mark=triangle,mark options={solid},forget plot]
  table[row sep=crcr]{%
0.00520833333333333	0.0112677200415071\\
0.00260416666666667	0.00160635757084918\\
0.00130208333333333	0.00275780939645404\\
0.000651041666666667	0.000790230272569913\\
};
\addplot [color=black,mark size=1.4pt,only marks,mark=square,mark options={solid},forget plot]
  table[row sep=crcr]{%
0.00520833333333333	0.0252181678838764\\
0.00260416666666667	0.00508806235554792\\
0.00130208333333333	0.001223080054045251\\
0.000651041666666667	0.0005344065218956203\\
};
\addplot [color=black,dashed,line width=0.5pt,forget plot]
  table[row sep=crcr]{%
0.00520833333333333	0.00520833333333333\\
0.00260416666666667	0.00260416666666667\\
0.00130208333333333	0.00130208333333333\\
0.000651041666666667	0.000651041666666667\\
};
\end{axis}

\begin{axis}[%
width=0.262\figW,
height=\figH,
at={(0.689\figW,0\figH)},
scale only axis,
xmode=log,
xmin=0.0001,
xmax=0.01,
xminorticks=true,
xlabel={$h$},
ymode=log,
ymin=0.01,
ymax=1,
ytick={0.01, 0.1, 1},
axis background/.style={fill=white},
title={$\frac{||u_x - u_{x,fine}||_\infty}{||u_{x,fine}||_\infty}$},
ticklabel style={font=\footnotesize},ylabel style={font=\footnotesize},xlabel style={font=\footnotesize},title style={font=\normalsize}
]
\addplot [color=black,mark size=1.4pt,only marks,mark=o,mark options={solid},forget plot]
  table[row sep=crcr]{%
0.00520833333333333	0.13983206346197\\
0.00260416666666667	0.0828759198830171\\
0.00130208333333333	0.0408818105440837\\
0.000651041666666667	0.013419125196732\\
};
\addplot [color=black,mark size=2pt,only marks,mark=diamond,mark options={solid},forget plot]
  table[row sep=crcr]{%
0.00520833333333333	0.152043230228994\\
0.00260416666666667	0.081291677474702\\
0.00130208333333333	0.0368793717461786\\
0.000651041666666667	0.0145946888878254\\
};
\addplot [color=black,mark size=2pt,only marks,mark=triangle,mark options={solid},forget plot]
  table[row sep=crcr]{%
0.00520833333333333	0.302955895142096\\
0.00260416666666667	0.130393313928447\\
0.00130208333333333	0.079696029891424\\
0.000651041666666667	0.0359314057412078\\
};
\addplot [color=black,mark size=1.4pt,only marks,mark=square,mark options={solid},forget plot]
  table[row sep=crcr]{%
0.00520833333333333	0.287630280259027\\
0.00260416666666667	0.155704230181416\\
0.00130208333333333	0.0781448850094658\\
0.000651041666666667	0.0348303832687004\\
};
\addplot [color=black,dashed,forget plot]
  table[row sep=crcr]{%
0.00520833333333333	0.21875\\
0.00260416666666667	0.109375\\
0.00130208333333333	0.0546875\\
0.000651041666666667	0.02734375\\
};
\end{axis}
\end{tikzpicture}%
    \caption[waste]{Errors in $\tilde{f}_x$, $C_D$, and $u_x$ versus grid spacing ($h$) for the cylinder in cross-flow problem. \protect\mycircle[none]\,: $\delta_h^{hat}$,  \protect\mydiamond[none]\,: $\delta_h^{3}$, \protect\mytriangle[none]\,: $\delta_h^{cos}$, \protect\mysquare[none]\,: $\delta_h^{G}$, \protect\dashedrule\,: first order convergence.}
    \label{fig:err_trans_cyl_Re200}
\end{figure}
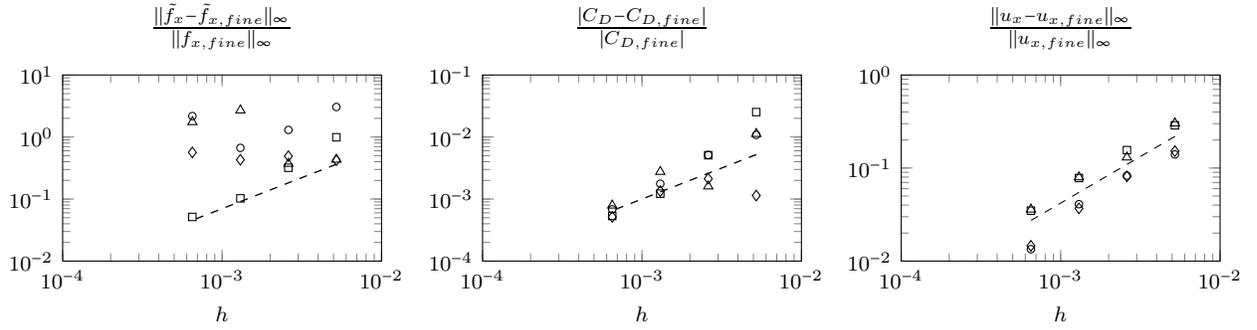

Figure \ref{fig:f_trans_cyl} demonstrates that the unphysical oscillations of the surface stresses $\tilde{f} = [\tilde{f}_x, \, \tilde{f}_y]^T$ are reduced for the smoother smeared delta functions. To demonstrate this quantitatively, we perform a convergence analysis by computing the infinity norm of the difference between $\tilde{f}_x$, $C_D$, and $u_x$ and the corresponding quantities obtained on a fine grid solution using $\delta_h^G$ with grid spacing $h = 4/3072\approx 0.001$. Similar results would be obtained using the $y$-components of $\tilde{f},$ $F$, and $u$. As with sections 2 and 4, $\delta_h^G$ yields surface stresses $\tilde{f}$ that converge to the fine-grid surface stress, but all smeared delta functions lead to convergent surface forces and velocities (see Figure \ref{fig:err_trans_cyl_Re200}). As with sections 2 and 4, the tildes are removed from force and velocity variables to emphasize that they are not affected by filtering. 

\begin{figure}[h!]
\setlength{\figH}{0.35\textwidth}
\setlength{\figW}{0.95\textwidth}
\centering
     \input{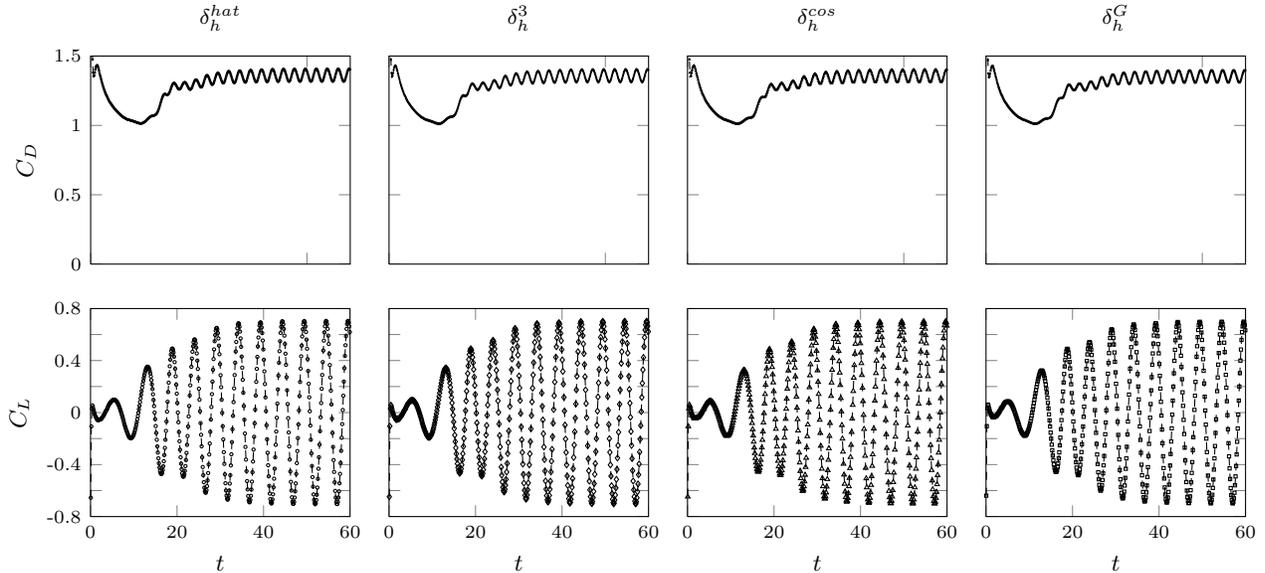}
    \caption[waste]{Coefficients of lift and drag on the cylinder versus time for the cylinder in cross-flow problem. The same grid spacing as in Figure \ref{fig:f_trans_cyl} was used.}
    \label{fig:F_trans_cyl}
\end{figure}

As seen in Figure \ref{fig:F_trans_cyl}, the present work faithfully replicates the well known periodic oscillations exhibited by $C_D$ and $C_L$ once the flow enters its limit cycle vortex shedding behavior. Table \ref{tab:freq} shows that the amplitude and dimensionless frequency ($St$) associated with these oscillations agree well with several previous experiments and simulations. This further demonstrates that accurate integral force values may be obtained irrespective of smeared delta function. Note by Figure \ref{fig:err_trans_cyl_Re200} that the integrated force is the same to within 10$^{-3}$ for all smeared delta functions considered. For simplicity we therefore only provide one value in Table \ref{tab:freq} with the understanding that it is representative of all smeared delta functions.  

\begin{table}[h!]
\centering
\begin{tabular}{| c | c | c | c |}
\hline
  & $St$ & $C_D$ & $C_L$ \\ \hline
  \cite{belov} & 0.193 & $1.19 \pm 0.042$ & $\pm 0.64$ \\ \hline
  \cite{liucylinder} & 0.192 & $1.31 \pm 0.049$ & $\pm 0.69$ \\ \hline
  \cite{laipeskin} & 0.190 & & \\ \hline
  \cite{roshko} & 0.19 &  & \\ \hline
  \cite{Taira:2007jl} & 0.196 & $1.35 \pm 0.048$ & $ \pm 0.68 $ \\ \hline
  Present & 0.198 & $1.35 \pm 0.046$ & $\pm 0.70$ \\ \hline
 \end{tabular}
 \caption{A comparison of of the dimensionless frequency ($St$) and amplitude of surface force oscillations}
\label{tab:freq}
\end{table}

\section{Conclusions}

The source of the inaccurate surface stresses and forces obtained by a class IB methods was identified: for any smeared delta function used, the equation for the surface stresses is an ill-posed integral equation of the first kind. As a result, the surface stresses computed from this equation have high frequency components that are erroneously amplified. We also demonstrated that the amplitude of the high frequency components of the physically correct surface stresses decreases as smoother smeared delta functions are used. Thus, for sufficiently smooth smeared delta functions, the incorrectly computed high frequency components may simply be filtered out to obtain accurate approximations to the actual stresses. We developed an efficient filtering technique that leads to better representations of the physical stresses than those obtained without filtering, and established that combining this filtering technique with an adequately smooth smeared delta function leads to surface stresses and forces that converge to the physical stresses and forces on the body. The filtering procedure is applied as a post-processing step, so it does not alter the convergent velocity field. We demonstrated the efficacy of the technique on two flow problems, flow in and around a rotating cylidner and flow over a circular cylinder, and demonstrate converged surface stresses in both cases.

\section{Acknowledgments}

This research was partially supported by a grant from the Jet Propulsion Laboratory (Grant No. 1492185). Many of the simulations were performed using the Extreme Science and Engineering Discovery Environment (XSEDE), which is supported by National Science Foundation grant number ACI-1053575. The first author gratefully acknowledges funding from the National Science Foundation Graduate Research Fellowship Program (Grant No. DGE--1144469). We thank Dr.\ Aaron Towne for insightful conversations about spectral decompositions of inverse operators, and Ms.\ Tess Saxton-Fox for her help in editing the manuscript. 

\section*{References}

\bibliography{force_bib}

\end{document}